\documentclass[
twocolumn,showpacs,prb,floatfix,
tightenlines,amsfonts,amssymb,eqsecnum
]{revtex4}

\usepackage{amsmath,bm,graphicx}

\begin{document}

\title{Quasiclassical theory of superconductivity:\\
interfering paths}

\author{M. Ozana}
\affiliation{Department of Theoretical Physics, 
Ume{\aa} University, 901 87 Ume{\aa}, Sweden}

\author{ A. Shelankov$*$}
\affiliation{Department of Theoretical Physics, 
Ume{\aa} University, 901 87 Ume{\aa}, Sweden}

 \date{\today}

\pacs{ 74.80Dm,74.80.-g,74.50.+r,74.20.Fg}

\begin{abstract}
We apply the method of two-point quasiclassical Green's function to
geometries where the trajectories include interfering paths and loops.
For a system of two superconducting layers separated by partially
transparent interface, corrections to the quasiclassical solutions for
the Green's function are explicitly found as well as the deviation
from the normalization condition.  
\end{abstract}

\maketitle

The method of quasiclassical Green's function has been long known as
the most efficient tool in the theory of superconductivity.  In the
quasiclassical theory, the superconductor is described by the matrix
Green's function which depends on the coordinate and the momentum as
well as the energy variable.  It obeys the Eilenberger equation
\cite{Eil68} or its nonstationary
generalisation\cite{LarOvc,Sch81,SerRai83}, the form of which is
similar to the Boltzmann kinetic equation, with the distribution
function replaced by a matrix.  Following the analogy between the
Eilenberger and Boltzmann equations, one may say that the Eilenberger
equation describes particles moving along classical trajectories, the
matrix nature of the ``distribution function'' accounting for the
internal quantum electron-hole degree of freedom of the quasiparticle
in a superconductor. As in the Boltzmann equation case, the notion of
a smooth trajectory remains meaningful in the presence of disorder:
Although individual impurities may lead to a large angle scattering,
the averaging with respect to the impurity positions restores the
spatial homogeneity, and the trajectories remain smooth. The presence
of disorder (leaving the localization effects out of the picture) is
seen only in a finite time that the particle lives on a given
trajectory.

In a mesoscopic superconductor, where during its life time the
particle may suffer multipe reflections on the boundary, or a bulk
sample with interfaces, the typical trajectory may be rather
complicated.  Compared with classical billiards, trajectories in a
superconductor become even more intricate due to the Andreev
reflection processes.  It has been recently argued in \cite{SheOza00}
that the derivation of the quasiclassical equation is heavily based on
the assumption that the trajectories are single-connected {\it i.e.},
there exists the unique path which connects any pair of points visited
by the particle: The one dimensional character of the trajectory
allows one to single out the phase factor due to the motion with the
Fermi momentum and derive an equation for a slowly varying envelope,
which is the Eilenberger equation.  Otherwise, if the particle has
alternative paths to go from one point to another, the quantum
interference between the paths takes place, which is sensitive to
variations of the path on the scale of the Fermi wave length
$\lambdabar_{F}$.  In this case, the Fermi wave length oscillations
cannot be disentangled from the behaviour at the larger spatial scales
of order of the coherence length.  Although interfering paths always
exist in any realistic situation (e.g. due multiple scattering on
impurity), the interference contribution is usually insignificant
since it vanishes after averaging with respect to the parameters of
the trajectories.  This is the reason why the quasiclassical
Eilenberger or Boltzmann equations, which completely ignore the
interference, work very well.

\begin{figure}[h] 
\includegraphics[height=0.2\textwidth]{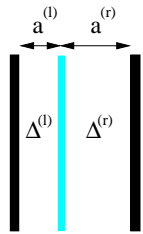}
\caption{
 Sandwich build of two
superconductor layers with the order parameters $\Delta^{(l,r)}$ and
thicknesses $a^{(l,r)}$ separated by  a partially transparent
interface. 
}
\label{2layer}
\end{figure}

\begin{figure}[h] 
\includegraphics[height=0.2\textwidth]{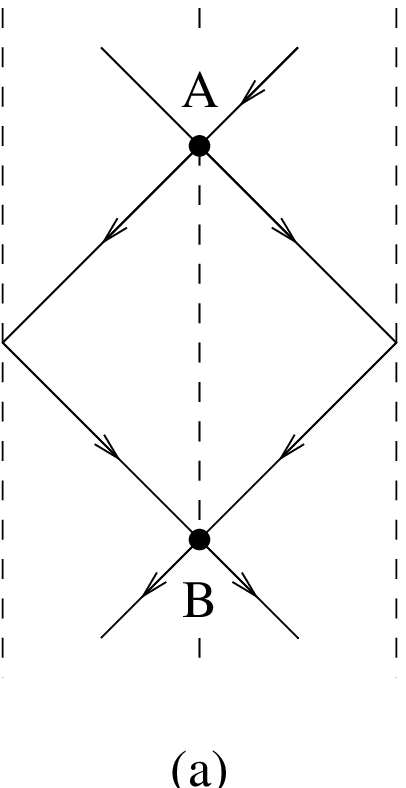}
\hspace*{.075\textwidth}
\includegraphics[height=0.2\textwidth]{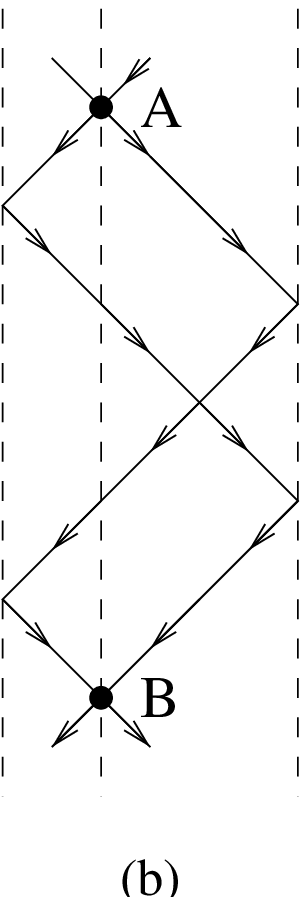}
\caption{
Traejectories in a double-layer system with a partially transparent
interface. In the ideal system, the reflections are specular.  (a)
When thicknesses of the layers are equal, the trajectories linked by
the collsion with the interface at the point A meet each other again
at the point B after the specular reflections by the boundaries.  (b)
For a general irrational relation between the layer thicknesses, the
trajectories meet each other meet again after two reflections.  The
lengths of the two interfering paths connecting the points A and B are
equal for any direction of the trajectory coming to the point A so
that the interference survives the momentum averaging.
}
\label{integrable} \end{figure}

However, there may be special reasons for the interference to survive
the averaging.  For instance, the interference of waves propagating
along two time reversed path is always constructive if time reversal
is a valid symmetry, and it leads to the well-known phenomena of weak
localization.  The interference may be robust in certain billiard-like
geometries where the alternative paths are created by reflections on
the boundaries or interfaces.  Here, one may invoke the analogy with a
classical billiard where intergrability is very sensitive to the
geometry.  As an example, consider an ideal two-layer system shown in
Fig. \ref{2layer}.  Trajectories for the case of equal thicknesses of
the layers is shown schematically in Fig. \ref{integrable} (a).  For
any direction of incidence, the two out-going waves created by the
collision with the specular interface combine together after
reflection by the outer walls.  The trajectory, that is, the set of
the points spanned by the particle, is not single-connected. The
lengths of the two alternative paths are the same so that their
interference is insensitive to the angular averaging and, therefore,
{\em may be} of importance. (Whether the interference indeed
influences observables requires an additional analysis.)  The simple
of the two recombining paths picture Fig. \ref{integrable} (a) exist
only if iwht a high precision the layers have equal thicknesses,
$a^{(l)}=a^{(r)}$.  However, the interference is possible for any
relation between $a^{(l)}$ and $a^{(r)}$: For the arbitrary parameter
$a^{(l)}/ a^{(r)}$, rational or irrational, the wave fragments
(partially) recombine after two collisions with the outer walls (see
Fig. \ref{integrable}(b)). It is only when the assumption about the
specular character of reflection is relaxed, it becomes statistically
impossible for the partial waves created by a collision with the
interface to meet each other ever again. In these conditions when the
alternative paths are absent, the trajectory becomes simply connected,
or tree-like in the terminology of \cite{SheOza00}.  Using the method
developed in \cite{SheOza00}, one is able to calculate the
quasiclassical Green's function on the tree-like trajectory.

The purpose of this paper is (i) to substantiate the conjecture
\cite{SheOza00} about the qualitative importance of the
single-connectivity of the trajectories in the conventional
quasiclassical theory; (ii) to suggest a method which allows one to
include the possibility of interfering paths in the quasiclassical
theory of superconductivity.  Being interested mainly in general
questions of the quasiclassical Green's function theory, we consider
for definiteness the simplest possible system where the interfering
path exist, namely, a sandwich shown in Fig.\ref{2layer}.  For the
specular case, this system is simple enough to solve the full Gor'kov
equation and thus to learn ``exact'' properties.  The exact solution
being known, one can judge on the validity of approximate schemes and
analyse the physical reason for their failure if it happens.  It has
been demonstrated numerically in \cite{OzaShe01} that the exact and
quasiclassical theory does give a different result in this case.  To
gain a better qualitative understanding of the numerical results of
\cite{OzaShe01} is the main goal of this study.

\begin{figure}
\centerline{\includegraphics[height=0.25\textwidth]{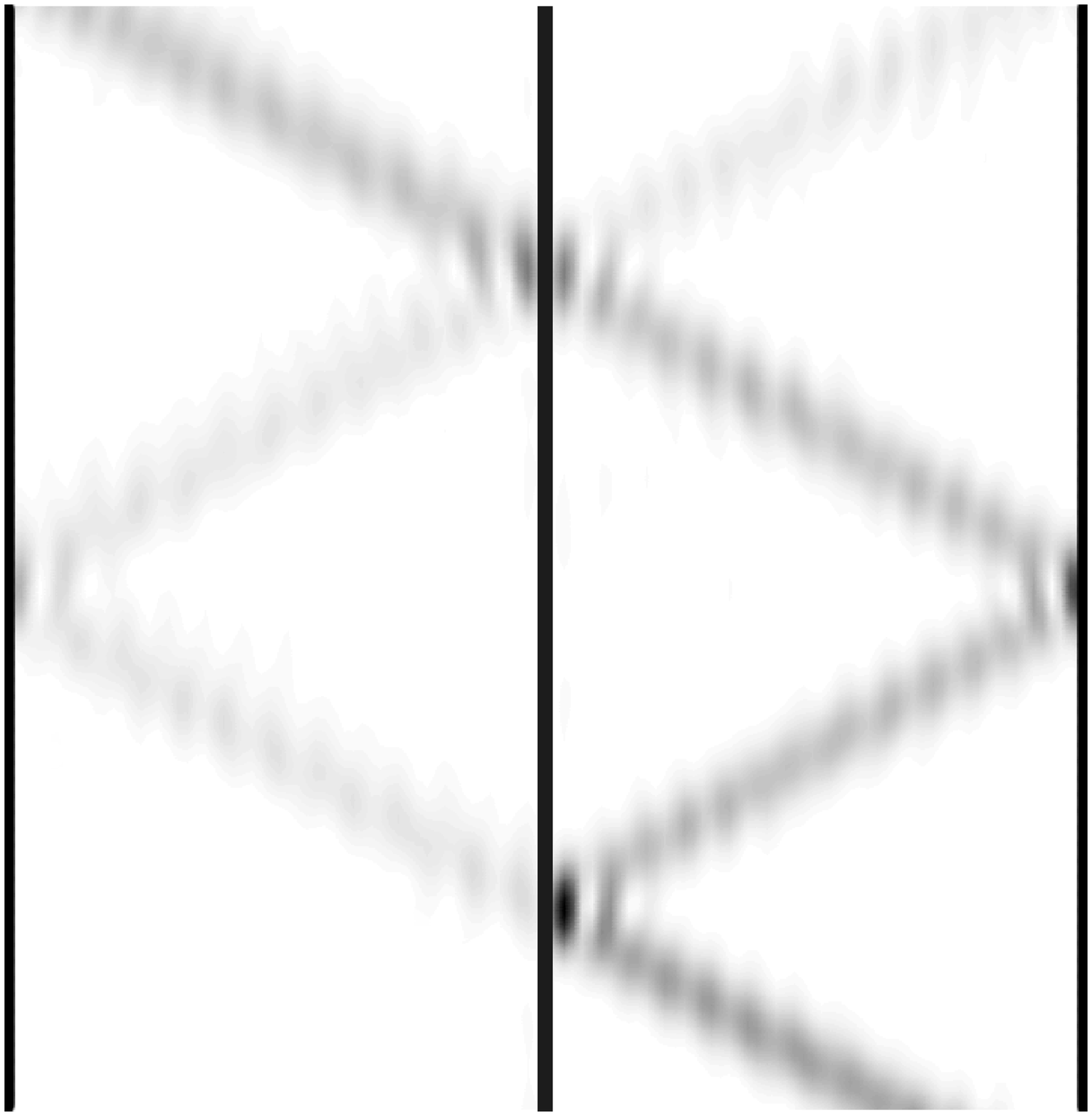}
 \hspace{0.051\textwidth}
\includegraphics[height=0.25\textwidth]{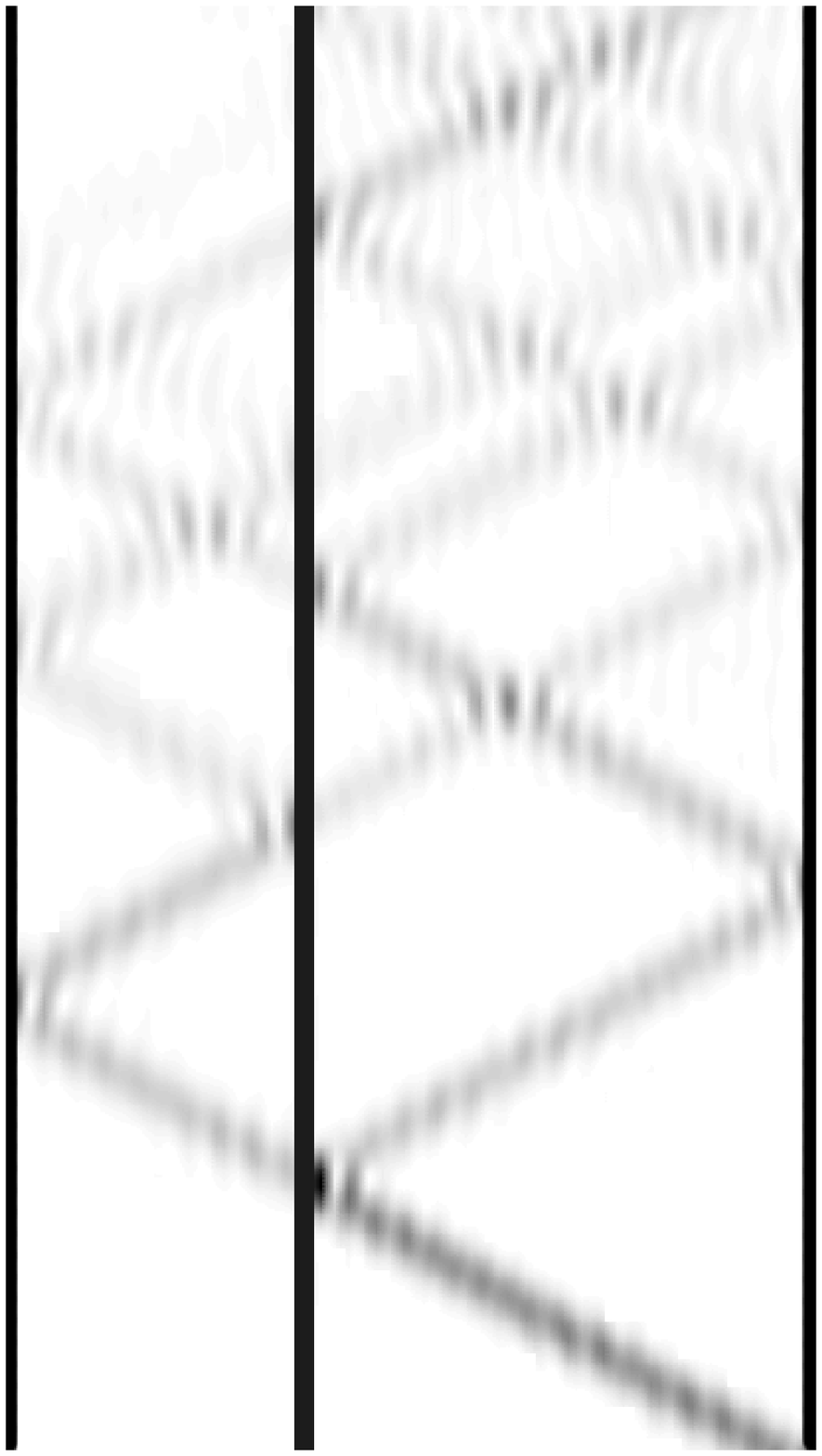}
}
\vspace{3ex}
\centerline{\hspace*{5ex}(a) \hspace*{25ex}(b)}
\caption{Trajectories in a sandwich is half-transparent interface. 
The intensity is proportional to the logarithm of the absolute value
of the exact normal metal Green function after Gaussian average in a
small region of the momentum parallel to the interface.  }
\label{coarse} 
\end{figure}

It is important to realize that the exact solution to the Gor'kov
equation must not be directly compared with the quasiclassical
solution. For instance, it is clear that the momentum resolved density
of states (DOS) found from the full Gor'kov equation for a layer with
specular walls, shows $\delta$-function peaks as a function of energy
or momentum due to the space quantisation in the layer. On the other
hand, the quasiclassical DOS is smooth as in an infinite layer. As
discussed in \cite{OzaShe01}, it is a coarse-grained Gor'kov Green's
function that should be compared with the quasiclassical results. The
coarse-grained function is generated by integration of the full
Green's function in a small region of $x-p$ space.  In particular, one
regains a classical picture of trajectories on the coarse-grained
level of description.  As an illustration of this point,
Fig. \ref{coarse} shows the picture of trajectories produced by the
integration of the exact normal metal Green's function $G(x,x';
\bm{p}_{||})\exp[i\bm{p}_{||}(\bm{r}- \bm{r}')]$ in a small region of
$\bm{p}_{||}$, $\bm{p}_{||}$ being the in-plane momentum.

The quasiclassical theory of superconductors with interfaces and walls
has been discussed in many papers.  Surfaces and interfaces violate
the condition of applicability of the quasiclassical theory, and their
presence is incorporated in the quasiclassical theory through a
boundary condition, again similar to the Boltzmann transport equation.
For the case of an impenetrable wall, the boundary condition has been
derived in Ref.\cite{BucRai79}. The matching condition for a partially
transparent specular interface has been derived in Ref.\cite{Zai84} as
well as \cite{SheOza00,Esc00}.  An isolated interface does not pose
severe problems for the quasiclassical theory of superconductivity.
The case of two and more interfaces turns out to be more difficult.
Analysing the solution to the full Gor'kov equation for a sandwich
made of two superconductors with a partially transparent interface,
some authors (see \cite{AshAoyHar89}, and \cite{Nag98}) have come to
the conclusion that the normalization of the Green's function is
violated and, therefore, the quasiclassical scheme fails in this
case. To a large extent, the present paper has been motivated by the
observation made in \cite{Nag98}.

The plan of the paper is the following. In Section \ref{two-point}, we
review the method of the 2-point Green's function
\cite{She85,SheOza00} on an infinite trajectory.  In Section
\ref{loop}, we derive quasiclassical Green's function for an isolated
loop, and discuss the normalisation condition. We show that the
normalization condition is violated before the coarse averaging, but
not after. In Section \ref{2loop}, we consider two coupled loops, and
show that the normalisation condition is not valid even on the
coarse-grained level. Finally, in Section \ref{sandwich}, we apply the
results to a superconducting two-layer and analyse applicability of
the quasiclassical theory.  The physical intepretation of the results
is given in Section\ref{disc}.

\section{ 2-point Quasiclassical Green's Functions}\label{two-point}

We start with a short review of the formulation of the quasiclassical
theory in terms of the two-point Green's function on classical
trajectories \cite{She85,SheOza00}. Here, we present  a novel
derivation of the equations in a form convenient for further
generalisations.

The main assumption of the quasiclassical theory of superconductivity
is the existence of two distinct spatial scales: the short one is the
Fermi wave length $\lambdabar_{F}= 1/ p_{F}$, $p_{F}$ being the Fermi
momentum, whereas the superconducting coherence length, $\xi_{0}$,
exemplifies the large scale. The quasiclassical theory is valid in the
zero approximation with respect to the small parameter
$\lambdabar_{F}/ \xi_{0}$ \cite{She80c,SerRai83}. In effect, the Fermi
wave length is put to zero and the orbital motion is essentially
classical, that is, the particle coordinate belongs to a chosen
classical trajectory. In this picture, the electron-hole mixing, which
is the essence of superconductivity, is a quantum internal degree of
freedom.

For a given momentum, the direction of velocity is opposite for the
electron and hole components of the wave functions, and the derivation
should start in a way that both direction on the trajectory enter on
equal footing. Consider propagation on a straight line and choose the
$x-$axis along the line. Denote the matrix $2\times2$ propagator
${\Bbb G}(x_{1},x_{2})$.  Throughout the paper, we deal with the
retarded Green's function of the Keldysh technique, $ {\mathbb
G}^{R}$; to simplify notation we omit the superscript $R$.  The
stationary Green's function satisfies the Gor'kov equation
\begin{equation}
(\varepsilon - {\Bbb H}) {\Bbb G}(x_{1},x_{2}) =  
\hat{1} \; \delta (x_{1}-x_{2})
\label{vec}
\end{equation}
where $\varepsilon = \varepsilon + i \delta $ is the energy variable
of the retarded Green's function, and the matrix operator ${\mathbb
H}$ reads
\begin{equation}
{\mathbb H} =
\left(
\begin{array}{cc}
\hat{\xi} & \Delta \\
\Delta^{*} & - \hat{\xi}
\end{array}
\right) - \Sigma
\label{wec}
\end{equation}
where $\Delta(x) $ is the superconducting order parameter, $\Sigma $
is the matrix (retarded) self-energy, and $\xi $ is the kinetic
energy,
\[
\hat{\xi } = 
\frac{\hat{p}_{x}^{2}}{2m} 
- \frac{p_{F}^{2}}{2m}
\quad,\quad 
\hat{p}_{x} = \frac{1}{i} \frac{d}{dx}
\;,
\]
with the Fermi energy subtracted.  For simplicity, we consider the
isotropic case in the absence of a magnetic field.  It is
straightforward to take into consideration the anisotropy of the order
parameter and the vector potential.

The boundary condition to Eq.~(\ref{vec}) is
\begin{equation}
{\mathbb G}(x_{1},x_{2}) \rightarrow 0  \quad,\quad |x_{1}-x_{2}| \rightarrow \infty  
\label{dfc}
\end{equation}
as it follows from the requirement of the analyticity of the Green's
function.

Considered as a function of the first argument $x_{1}$, the Green's
function ${\mathbb G}$ in Eq.~(\ref{vec}) describes the wave
propagating from the point $x_{2}$.  If the energy variable
$\varepsilon $ is small in comparison with the Fermi energy
$p_{F}^2/2m$, the wave has two components with the momenta close to
$+p_{F}$ or $- p_{F}$.  In the quasiclassical theory, these components
are ascribed to two trajectories with the opposite momentum, and the
corresponding trajectory propagators obey separate inhomogeneous
equations.  To carry out the separation, we use the following
procedure.

The $\delta -$function inhomogeneity in Eq.~(\ref{vec}), that is the
source of the wave, is equivalent to the following conditions at the
points $x_{1}=x_{2}\pm 0$
\begin{eqnarray}
{\mathbb G}(x_{1}, x_{2}) 
\rule[-1.5ex]{.051ex}{4ex}
_{\hspace*{.5ex}x_{1}= x_{2} -0}
^{\hspace*{.5ex}x_{1}= x_{2} +0}    
& =  & 0              \label{4ec}\\   
\frac{1}{2m} \frac{d}{d x_{1}}
{\mathbb G}(x_{1}, x_{2}) \rule[-1.5ex]{.051ex}{4ex}
_{\hspace*{.5ex}x_{1}= x_{2} -0}
^{\hspace*{.5ex}x_{1}= x_{2} +0}
    & =  & \tau_{z}           \;.     \label{4ec2}
\end{eqnarray}

Defining the differential  operators $D^{\pm}_{x}$,
\[
D^{\pm}_{x}= \frac{1}{2} \pm \frac{1}{ 2i\, p_{F} } \; \frac{d}{dx}\;,
\]
one rewrites  
Eqs.~(\ref{4ec}) and   (\ref{4ec2}), 
as the following two inhomogeneous equations,
\[
i v_{F} D_{x_{1}}^{\pm}
{\mathbb G}(x_{1}, x_{2}) 
\rule[-1.5ex]{.051ex}{4ex}
_{\hspace*{.5ex}x_{1}= x_{2} -0}
^{\hspace*{.5ex}x_{1}= x_{2} +0}
=  \hat{1}  \;.
\]

These conditions can be satisfied by the following decomposition:
\begin{equation}
{\mathbb G}(x_{1},x_{2}) = \sum\limits_{\sigma =\pm} {\mathbb G}^{\sigma }(x_{1},x_{2}) \;,
\label{7ec}
\end{equation}
where the partial propagators ${\mathbb G}^{\pm}$ obey the {\em
homogeneous} linear equation Eq.~(\ref{vec}) at $x_{1}\neq x_{2}$ and
the {\em inhomogeneous} conditions
\begin{equation}
i v_{F} D_{x_{1}}^{\sigma' }
{\mathbb G}^{\sigma }(x_{1}, x_{2}) \tau_{z}
\,
\rule[-1.5ex]{.1ex}{3ex}
_{\,x_{1}=x_{2}- \sigma \cdot 0}
^{\,x_{1}=x_{2}+ \sigma \cdot 0}
= \delta_{\sigma \sigma '}\hat{1}
\quad,\quad 
\sigma , \sigma ' = \pm
\; .
\label{8ec}
\end{equation}
which replace the $\delta $-function source in the r.h.s. of
Eq.~(\ref{vec}).

Now, one observes that
${\mathbb G}^{+}$ and ${\mathbb G}^{-}$ correspond to the opposite
direction of the Fermi momentum
{\it i.e.}  they can be presented as
\begin{equation}
i v_{F} \;{\mathbb G}^{\pm}(x_{1},x_{2}) \tau_{z}  =
g^{\pm}( x_{1},x_{2}) \; e^{\pm i p_{F}(x_{1}-x_{2})}\;,
\label{9ec}
\end{equation}
where $g^{\pm}( x_{1},x_{2})$ is a slowly varying function if the
energy $\varepsilon $ is small in comparison with the Fermi energy;
the factor with the Fermi velocity $v_{F}= p_{F}/m$ and the Pauli
matrix $\tau_{z} $ are introduced for future convenience.

Indeed, for slowly varying $g^{\pm}$,
\begin{equation}
{1\over i } {d \over{dx_{1}}} 
g^{\pm}(x_{1},x_{2}) e^{\pm i p_{F}(x_{1}-x_{2})} 
\approx
\pm p_{F}\;g^{\pm}(x_{1},x_{2}) e^{\pm i p_{F}(x_{1}-x_{2})}
\;.
\label{zec}
\end{equation}
Assuming that typically $g^{\pm}(x_{1},x_{2})$  vary on the coherence
length $\xi_{0}$,
the corrections to this relation are of order of $\lambdabar_{F}/
\xi_{0}$. Small quantities of this order are consistently neglected in
the quasiclassical theory.  Note that this formula correspond to the Andreev
approximation \cite{And64} where the derivatives of 
quasiclassical envelope functions
are always neglected if they stand next to
$p_{F}$.

\begin{figure}
\includegraphics[height=0.3\textwidth]{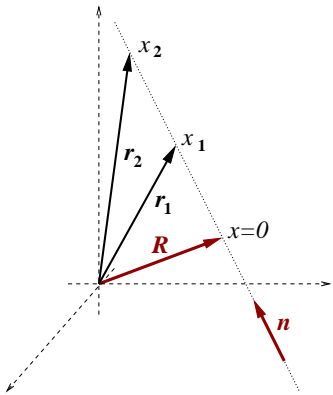}
\caption{ Classical trajectory.  
When no interfaces or boundaries are present the trajectory is a
straight line. It can be specified by an initial point $\bm{R}$ and
the unit vector $\bm{n}$, which shows the direction of the trajectory.
Introducing the coordinate along the trajectory $x$, any point
$\bm{r}$ on the trajectory can be written as $\bm{r} = \bm{R} + x
\bm{n}$.} 
\label{trajectory}
\end{figure}

With the accuracy Eq.~(\ref{zec}) holds, $D_{x}^{\pm}$ are projectors,
{\it i.e.},
\[
D_{x_{1}}^{\sigma}
\sum\limits_{\sigma' =\pm} 
{\mathbb G}^{\sigma' }(x_{1},x_{2}) =
{\mathbb G}^{\sigma }(x_{1},x_{2}) \;,
\]
The projecting property guarantees that one of the conditions in
Eq.~(\ref{8ec}), namely, corresponding to $\sigma '= - \sigma $, is
automatically satisfied. The remaining conditions read
\begin{eqnarray}
 g^{+}(x_{1}, x_{2})\rule[-1.5ex]{.051ex}{4ex}
_{\hspace*{.5ex}x_{1}= x_{2} -0}
^{\hspace*{.5ex}x_{1}= x_{2} +0}
& =  & \hat{1}      \; ,
\rule[-2.5ex]{0ex}{0ex}
        \label{bfc}\\   
 g^{-}(x_{1}, x_{2})\rule[-1.5ex]{.051ex}{4ex}
_{\hspace*{.5ex}x_{1}= x_{2} +0}
^{\hspace*{.5ex}x_{1}= x_{2} -0}
    & =  & \hat{1}      \;.
        \label{bfc2}   
\end{eqnarray}
It is clear that $g^{+}$ and $g^{-}$ are associated with the Fermi
surface points $+ p_{F} \bm{n}$ and $- p_{F} \bm{n}$, respectively,
$\bm{n}$ being the orientation of the $x$-axis.  Spanning different
orientation of the $x-$axis, one defines function $g(x_{1},x_{2}|
\bm{n},\bm{R})$ where $ \bm{n}$ and $\bm{R}$ are the parameters of the
trajectory, and $x_{1,2}$ are the coordinates on the trajectory see
Fig.\ref{trajectory}. Then information on the direction of the
momentum is contained in $\bm{n}$, and the superscript $\pm$ in
$g^{\pm}$, being redundant, can be omitted.

Plugging the ansatz Eq.~(\ref{9ec}) into Eq.~(\ref{vec}), and using
the Andreev approximation
\[
e^{- i p_{F}x} \;
\hat{\xi }\;
e^{ i p_{F}x}\approx -i v_{F} {\partial\over{\partial x}}
\]
one 
obtains 
the first
order differential equation
for $g$. 

The 2-point Green's function
$\hat{g}_{\varepsilon}(x_{1},x_{2}|\bm{n},\bm{R})$ obeys the following
set of equations \cite{She85}:
\begin{widetext}
\begin{subequations}
\label{vra} 
\begin{eqnarray}
\left( 
i v_{F} {\partial \over \partial x_{1}} - \Phi(\bm{r}_{1}) 
+\hat{H}_{\varepsilon,\bm{n}}(\bm{r}_{1})\right)
\hat{g}_{\varepsilon}(x_{1},x_{2}|\bm{n},\bm{R})
&=&
i v_{F}  \delta(x_{1} - x_{2}) 
,\;
\label{vraa}\\   
\hat{g}_{\varepsilon}(x_{1},x_{2}|\bm{n},\bm{R})
\left( 
-i v_{F} {\partial \over \partial x_{2}} -\Phi(\bm{r}_{2}) + 
\hat{H}_{\varepsilon,\bm{n}}(\bm{r}_{2})\right)
&=&
i v_{F}  \delta(x_{1} - x_{2}) 
,\;
 \label{vrab}
\end{eqnarray}
\end{subequations}
\end{widetext}
where the derivative in Eq.~(\ref{vrab}) operates backwards.  The
$\delta $-functions in the r.h.s.  generates the discontinuity
required by Eqs.~(\ref{bfc}).

In Eq.~(\ref{vra}), $\Phi $ is the gauge invariant scalar potential,
and $\hat{H}_{\varepsilon, \bm{n}}$ is $2\times 2$ traceless matrix,
\[
\hat{H}_{\varepsilon, \bm{n}}=  
 \hat{h }_{\varepsilon, \bm{n}}
- \hat{\Sigma }_{\varepsilon, \bm{n}}
\;\; ,  
\]
\[
\hat{h}_{\varepsilon, \bm{n}}=
\left(
\begin{array}{lr}
\varepsilon - \bm{v}_{F} \cdot \bm{p}_{s}&  -\Delta_{\bm{n}}\\
\Delta^{*}_{\bm{n}}& -\varepsilon + \bm{v}_{F} \cdot \bm{p}_{s}
\end{array}
\right) 
\;\; , \;\;  \bm{v}_{F}= v \bm{n} \; ,
\]
where $\Delta_{\bm{n}}$ is the order parameter (which may depend on
the direction $\bm{n}$), and $\bm{p}_{s}= - {e\over c}\bm{A}$,
$\bm{A}$ being the vector potential, and $\hat{\Sigma}$ is built of
the impurity self-energy as well as the part of the electron-phonon
self-energy not included to the self-consistent field $\Delta $.
Below, we omit $\bm{R}, \bm{n}$ and $\varepsilon$ for brevity.

The boundary condition to Eq.~(\ref{vra}) on an infinite trajectory
from Eq.~(\ref{dfc}),
\begin{equation}
\hat{g}(x_{1} , x_{2}) \rightarrow 0 
\quad  \text{ for } \quad
| x_{1} - x_{2} | \rightarrow \infty \; .
\label{Mswb}
\end{equation}

The equations of motion Eq.~(\ref{vra}) can be solved  
by factorizing 
the Green's function \cite{She85,SheOza00}.  For this, one finds
linear independent two-component
solutions $\phi_{\pm}(x)= {u_{\pm}\choose v_{\pm}}$ to the equation
\begin{equation}
\left(
i v_{F} {\partial \over \partial x} +
\hat{H}\right) \phi_{\pm}
=0 \;,
\quad,\quad 
\phi_{\pm}(x) \rightarrow 0 \quad,\quad x \rightarrow \pm \infty 
\label{yfc}
\end{equation}
where $H$ is in the operator on the l.h.s. of Eqs.~(\ref{vra}); This
equation is essentially the same as the one first derived by Andreev
in the theory of reflection on the NS interface.  The ratio
$u_{\pm}/v_{\pm}$ obeys the Riccati equation \cite{Nag98,SchMak95},
which is most convenient for numerics. Known solution to the Riccati
equation for $u_{\pm}/v_{\pm}$, one readily finds full $\phi_{\pm}$ as
explained in \cite{SheOza00}.

The solutions to the linear equation Eq.~(\ref{yfc}) can always be
normalized as
\begin{equation}
\bar{\phi }_{-}\phi _{+} =1
\label{zfc}
\end{equation}
where here and below the bar on the top of a 2-component column,
$\psi$, or a matrix $X$, has the following meaning
\begin{equation}
\bar{\psi }= \psi^{T} \tau_{y} {1\over i}
\quad,\quad 
\bar{X} = \tau_{y} X^{T} \tau_{y} 
\label{2fc}
\end{equation}
where $T$ stands for transposition and $\tau_{y} $ is the Pauli
matrix. Taking advantage of the property $H = - \bar{H}$, one
can prove that the normalization in Eq.~(\ref{zfc}) is generally
compatible with Eq.~(\ref{yfc}).

Denote also
\[
\phi_{\nu }^{\ddagger} = \nu \bar{\phi}_{-\nu } \quad,\quad  \nu = \pm
\; .
\]
Then,
\begin{equation}
\phi ^{\ddagger}_{\nu } \phi_{\nu '} = \delta_{\nu \nu '}
\quad,\quad 
\sum\limits_{\nu } \phi_{\nu}\phi ^{\ddagger}_{\nu }
= \hat{1}
\label{4fc}
\end{equation}
as immediately follows from the normalization Eq.~(\ref{zfc}) and the
definitions in Eq.~(\ref{2fc}).

Now, the solution to  Eq.~(\ref{vra}) reads 
\begin{equation}
\hat{g}(x_{1},x_{2})=
\left\{
\begin{array}{rcr}
\phi_{+}(x_{1})\; \phi_{+}^{\ddagger}(x_{2})   &\;\;,\;\;& x_{1}> x_{2}  \;;\\
-\;\phi_{-}(x_{1})\; \phi_{-}^{\ddagger}(x_{2}) 
  &\;\;,\;\;& x_{1}< x_{2}      \;.
\end{array}
\right.
\label{5fc}
\end{equation}

In the standard formulation of the quasiclassical technique after
Eilenberger, one deals with the 1-point Green's function $\hat{g}(x)$,
which is related to the 2-point Green's function as \cite{She85}
\begin{equation}
\hat{g}(x) = 
\hat{g}_{+}(x)
+
\hat{g}_{-}(x)
\label{efc}
\end{equation}
where 
\[
\hat{g}_{\pm}(x)
=
\hat{g}(x_{1}= x\pm0\;,\;x_{2}=x\mp 0)
\]
Combining Eqs.~(\ref{vraa}),  and (\ref{vrab}), one re-derives 
the well-known Eilenberger equation \cite{Eil68},
\[
i \bm{v}_{F} \bm{\cdot \nabla }\hat{g} + 
\hat{H}\hat{g} - \hat{g}\hat{H} =0 \; ,
\]
for the matrix function $\hat{g}$.  The expressions for observables
like the charge current and the density of states are presented in
Section \ref{obser}.

In the factorized representation Eq.~(\ref{5fc}), the normalization
condition for the Eilenberger function Eq.~(\ref{efc}),
\begin{equation}
\left(\hat{g} \right)^2 = \hat{1} 
\label{hfc}
\end{equation} 
immediately follows from the orthonormality relations in Eq.~(\ref{4fc}).

To trace the general features needed for the validity of
Eq.~(\ref{hfc}), we first note that
\[
\hat{g}_{+}(x)
-
\hat{g}_{-}(x)
= \hat{1}
\; 
\]
as is dictated by the $\delta-$function in the r.h.s. of
Eq.~(\ref{vra}).  Taking square of this equation and Eq.~(\ref{efc}),
one gets the normalization condition Eq.~(\ref{hfc}) provided
\[
g_{+}(x)g_{-}(x) = 
g_{-}(x)g_{+}(x) = 0 \;. 
\]
The proof of these relations presented in Ref. \cite{She85},
essentially uses the asymptotics in Eq.~(\ref{Mswb}), which makes
sense only for an infinite trajectory.  Besides, the proof tacitly
presumes that the quasiclassical equations of motion Eq.~(\ref{vra})
are applicable everywhere on the trajectory. The latter assumption is
not valid if the trajectory under consideration is crossed by another
trajectory -- the trajectory crossing happens, for instance, on
partially transparent interfaces. As shown in Ref. \cite{SheOza00},
the trajectory crossings do not influence the normalization provided
the crossings do not change the simple one-dimensional topology of the
trajectories. It is the purpose of the present paper to analyse more
general cases.

After the review of the two-point quasiclassical Green's function
method, we will extend the method to the case when the classical
trajectories have more complicated topology. In the standard theory,
the Fermi wave length scale does not enter the theory after the
factorization in Eq.~(\ref{9ec}) has been carried out. As we will see
later, this disentanglement of the different scales can not be
performed in the case of more complicated geometry.

\section{Green's function on a loop}\label{loop}

The theory reviewed in the previous section refers to the situation
when the trajectory is a straight line.  One can easily generalise the
derivation to the case when the particle moves in a smooth
quasiclassical potential and its trajectories are not straight
lines. If the trajectory corresponds to the infinite motion, no
important changes in the theory occur: one can use Eq.~(\ref{vra})
with the understanding that $x$ is the coordinate along the path. If
the orbit is closed, the situation is very different. Physically, the
obvious difference is that if the particle moves along a trajectory it
eventually comes again to the starting point; then, statements like
$x_{1}>x_{2}$ do not have unequivocal sense.  Also, the boundary
condition in Eq.~(\ref{Mswb}), without which the equation of motion
Eq.~(\ref{vra}) do not have a unique solution, becomes meaningless.

Of course, in a clean homogeneous metal one can hardly imagine any
smooth loop-like trajectory because any potential is screened by
mobile charges.  Actually, we have in mind to consider loops created
as a result of multiple reflections by interfaces. However, for the
sake of presentation, it is convenient to us to analyse first a smooth
closed trajectory. Besides, the problem is of general interest.

\begin{figure}[htbp]
\includegraphics[height=0.2\textwidth]{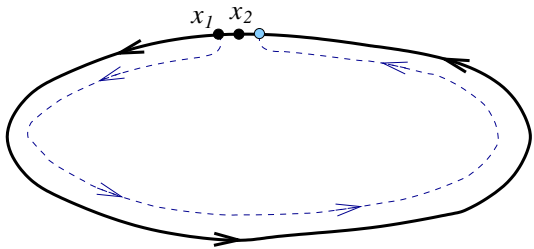}
\caption{A loop.  The Green's function is a function of the two
coordinates along the loop $x_{1}$ and $x_{2}$.
As shown in the picture, advancing gradually the first argument from the
position $x_{1}= x_{2} +0$ in the direction of its increase, one 
approaches the point marked by the dashed circle which locally correponds
to the coordinate   $x_{1}=x_{2}-0$. 
}
\label{looop}
\end{figure}

 Consider a closed loop trajectory shown in Fig.\ref{looop}.  We
accept it without derivation that Eq.~(\ref{vra}) may serve as the
equation of motion for the two point trajectory Green's function. What
is needed is the boundary condition replacing Eq.~(\ref{Mswb}).

First, we recall that the full propagator is given by ${\mathbb G}$ in
Eq.~(\ref{9ec}),
\begin{equation}
i v_{F} \;{\mathbb G}(x_{1},x_{2}) \tau_{z}  =
g( x_{1},x_{2}) \; e^{ i p_{F}(x_{1}-x_{2})}\;,
\label{kfc}
\end{equation}
omitting the redundant superscripts $\pm$.  Consider the Green's
function at nearly coinciding arguments, ${\mathbb
G}(x_{1}=~x_{2}+0,x_{2})$. Using equation of motion, one can find
${\Bbb G}(x_{1},x_{2})$ advancing $x_{1}$ in the direction of arrows
in Fig.\ref{looop}. Eventually, one gets again close to $x_{2}$ when
$x_{1}$ reaches the value $x_{2} + {\cal L} -0$, ${\cal L}$ being the
circumference of the loop.  On the other hand, the point $x_{1}=x_{2}
+ {\cal L} -0$ is the same physical point as $x_{1}=x_{2}- 0$.  From
the requirement that the Green's function is uniquely defined, one
gets the following boundary condition
\[
{\mathbb G}(x_{1}= x_{2}+ {\cal L},x_{2}) =
{\mathbb G}(x_{1}= x_{2}-0,x_{2}) \; .
\]

From here one gets the following boundary condition for the
quasiclassical Green's function 
\begin{equation}
e^{i p_{F} {\cal L}} 
\hat{g}(x_{2} + {\cal L}, x_{2})=
\hat{g}(x_{2}-0, x_{2} )
\label{nfc}
\end{equation}
where $\hat{g}(x_{2} + {\cal L}, x_{2})$ is the quasiclassical
Green's function  continued  
with the help of   Eq.~(\ref{vra})
from  $x_{1}= x_{2}+0$ along the loop of the length ${\cal L}$ to
the point $x_{1}= x_{2}+{\cal L} $.

As the next step, we derive formally exact expression for the 1-point
quasiclassical Green's function via the quasiclassical evolution
operator $U(x,x')$. The latter solves the homogeneous part of
Eq.~(\ref{vra}),
\begin{equation}
\left(
i v_{F} {\partial \over \partial x} +
\hat{H}
\right)
U(x,x')=0 \;,
\label{pfc}
\end{equation}
with the initial condition $U(x,x)=1$.  A formal solution to
Eq.~(\ref{pfc}) reads
\begin{equation}
U(x, x') = \exp\left[ 
{i\over v_{F}} \int\limits_{x'}^{x} \; dx
\;\hat{H}
\right] 
\label{6fc}
\end{equation}
where $x-$ordering of the matrices $\hat{H}(x)$ is implied.
Properties of $U$ has been discussed in many papers, see
e.g. \cite{Nag98}. In particular, $\det U =1$, and
\begin{equation}
U(x,x') \bar{U}(x,x') = \hat{1}
\label{qfc}
\end{equation}
The expression for the evolution operator in the Riccati equation
technique is presented in Section \ref{U} (see also \cite{SheOza00}).

Denote
\[
U_{x} = U(x+ {\cal L}, x) \quad,\quad {\cal U}_{x}= e^{i p_{F}{\cal
L}} U_{x} \; .
\]
Then $\hat{g}(x + {\cal L}, x)=
U_{x}
\hat{g}(x+0, x )$ and the boundary condition Eq.~(\ref{nfc})
acquires the form
\[
{\cal U}_{x}
\hat{g}(x+0, x )
=
\hat{g}(x-0, x )
\]
This equation together with 
\[
\hat{g}(x+0, x )
- 
\hat{g}(x-0, x ) = \hat{1}
\]
allows one to obtain the expression for the one-point Green's
functions via the evolution operator:
\begin{subequations}
\label{xfc}
\begin{eqnarray}
\hat{g}_{+}(x)
& =  & {1\over 1- {\cal U}_{x}}             \label{xfca}\\   
\hat{g}_{-}(x)     
& =  &  {{\cal U}_{x}\over 1- {\cal U}_{x}}
\nonumber 
\\
\hat{g}(x)& =  &  {{1 + \cal U}_{x}\over 1- {\cal U}_{x}}             \label{xfcc}
\end{eqnarray}
\end{subequations}
For illustration,  we consider some simple applications of this result.

\subsubsection{Homogeneous loop}

Suppose the ``Hamiltonian'' $\hat{H}$ in Eq.~(\ref{pfc}) does not
depend on the coordinate. The evolution operator $U_{x}$ can be easily
found from Eq.~(\ref{6fc}). For this, we present $\hat{H}$ as
\[
\hat{H} = \xi \;\psi_{+} \psi_{+}^{\ddagger} - \xi \;\psi_{-}\psi_{-}^{\ddagger}
\]
where $\psi_{\pm}$ are the eigenfunction of $\hat{H}$,
\begin{equation}
\hat{H} \psi_{\pm} =\pm \;\xi\; \psi_{\pm} 
\quad,\quad  \Im \,\xi > 0
\label{9fc}
\end{equation}
normalized  in accordance with Eq.~(\ref{4fc}).
From Eq.~(\ref{6fc}), the loop evolution operator reads
\begin{equation}
U_{x} =  
u \; \psi_{+}\psi_{+}^{\ddagger}
 +  {1\over u}\;\psi_{-}\psi_{-}^{\ddagger}
 \;,
\label{7fc}
\end{equation}
independent on $x$. 
Parameter $u$, 
\begin{equation}
u = e^{{i \xi {\cal L}\over v_{F} } }
\label{2hc}
\end{equation}
shows the variation of the phase of the wave and its absolute value
upon propagation around the loop.  The modulus of $u$, $|u|=
\exp[-{\cal L}\; \Im \xi / v_{F}] <1$, is less than unity.

The one-point Green's function Eq.~(\ref{xfcc}) reads
\begin{equation}
\hat{g} = 
{1 +  u\,e^{i p_{F}{\cal L}} \over 1 - u\,e^{i p_{F}{\cal L}}}
\;\psi_{+}\psi_{+}^{\ddagger}
-
 {1 + u\,e^{-i p_{F}{\cal L}} \over 1- u\,e^{-i p_{F}{\cal L}}}
\;\psi_{-}\psi_{-}^{\ddagger}
\; .
\label{agc}
\end{equation}

This result is rather transparent physically. The Green's function has
a contribution from (quasi)electrons and (quasi) holes given by the
first and second terms, respectively.  The (retarded) Green's function
is built of waves decaying in the direction of propagation, and one
can use attenuation of the wave, $|u|$, as a meter which shows the
length of the path the particle has travelled.  Expanding the Green's
function in Eq.~(\ref{agc}) with respect to $u$,
\begin{eqnarray}
\hat{g}     & =  &  
\psi_{+}\psi_{+}^{\ddagger} \left(1 + u\,e^{i p_{F}{\cal L}} +
\left(u\,e^{i p_{F}{\cal L}} \right)^2 + \ldots\right)
            \nonumber\\   
&-&     
 \psi_{-}\psi_{-}^{\ddagger}
\left(1 +u\,e^{-i p_{F}{\cal L}} +
\left(u\,e^{-i p_{F}{\cal L}} \right)^2 + \ldots\right)
\; .
\nonumber 
\end{eqnarray} 
one gets information on the paths contributing the propagator and
their length.  Looking at the first line, one sees that the Green's
function has a contribution which corresponds to the electron-like
quasiparticle $\psi_{+}$ circling the loop in the positive direction
(where the phase in Fermi momentum factor {\em increases}); the paths
have the winding numbers $0,1,2\ldots$ and the lengths $0, {\cal L},
2{\cal L},\ldots$, respectively. The second line shows the
contribution due to holes: Decaying in the direction of propagation,
they circle the loop in the negative direction as can be read off the
Fermi momentum phase factor \cite{gA}.  In the limit of an infinite
loop, when $u \rightarrow 0$, one restores the well-know result for a
homogeneous superconductor.

The electron and hole parts in Eq.~(\ref{agc}) have poles at the
energies where the denominators have zeroth. In the ballistic case,
when the non-Hermitian part of $H$ is only due to the infinitesimal
$\delta$ in $\varepsilon + i \delta $, the poles give the bound states
which correspond to the usual Bohr-Sommerfeld quantisation for closed
orbits.  Impurities, generating the imaginary part of the self-energy,
reduce $u$ and broaden the bound states. For quantities averaged with
respect to disorder, the closed topology of the orbit is seen only if
the orbit is shorter than the mean free path.  At Matsubara
frequencies $\varepsilon = i \omega_{n}$, the orbit must be shorter
than the coherence length $v_{F}/ \omega_{n}$.

 A qualitative difference between the cases of a loop
and an infinite trajectory, is clearly seen in the expansion of
$\hat{g}_{+}$ Eq.~(\ref{xfca}),
\begin{eqnarray}
\hat{g}_{+}     & =  &  
\psi_{+}\psi_{+}^{\ddagger} \left(1 + u\,e^{i p_{F}{\cal L}} +
\left(u\,e^{i p_{F}{\cal L}} \right)^2 + \ldots\right)
            \nonumber\\   
&-&     
 \psi_{-}\psi_{-}^{\ddagger}
\left(u\,e^{-i p_{F}{\cal L}} +
\left(u\,e^{-i p_{F}{\cal L}} \right)^2 + \ldots\right)
\; .
             \label{fgc}  
\end{eqnarray} 
One sees that, unlike the case of an infinite trajectory, the hole
contributes to $\hat{g}_{+}$. For this, the hole must circle the loop
at least once.  The reason for the difference between electron and
hole is physically clear.  Indeed, by definition, $g_{+}(x)$ is the
propagator $g(x+0, x)$ from the source at the point $x$ to the
neighbouring point $x +0$.  The source ({\it i.e.}  the r.h.s of
Eqs.~(\ref{vraa}) creates electron which propagates in the positive
direction, and the hole moving in the negative direction, nominally,
away from the point $x+0$. However, having made a full turn(s) around
the loop, the hole reaches the point $x+0$, and, therefore, does
contribute to $\hat{g}_{+}$.  Ultimately, these ``wrong'' processes
where the excitation effectively moves in the wrong direction, lead to
the violation of the normalization condition for the quasiclassical
Green's function.

\subsubsection{Normalization condition}

To check the normalization condition, we note that $\hat{g}^2 -1$ can
be conveniently presented as
\[
\hat{g}^2 -1 = 4\; \hat{g}_{+}\hat{g}_{-}
\]
From Eq.~(\ref{xfc}),
$\hat{g}_{+}\hat{g}_{-}= {\cal U}_{x}/(1 - {\cal U}_{x})^2$, or
in the eigenfunction representation, 
\begin{equation}
\hat{g}_{+}\hat{g}_{-} =
{u\,e^{i p_{F}{\cal L}} \over (1 - u\,e^{i p_{F}{\cal L}})^2}
\;\psi_{+}\psi_{+}^{\ddagger}
+
 {u\,e^{-i p_{F}{\cal L}} \over (1- u\,e^{-i p_{F}{\cal L}})^2}
\;\psi_{-}\psi_{-}^{\ddagger}
\; .
\label{bgc}
\end{equation}
One can see that the orthogonality and normalization condition do {\em
not} hold for the Green's function on a loop.  Of course, this is not
informative from the point of view of checking the validity of the
quasiclassical theory which deals with coarse-grained objects.  Note,
however, the r.h.s. of Eq.~(\ref{bgc}) vanishes when $u\rightarrow 0$,
restoring the normalization Eq.~(\ref{hfc}) in the limit of an
infinite loop.

\subsubsection{Coarse-grained average}

Observables like the local density of states (DOS) calculated with the
help of the Green's function in Eq.~(\ref{xfcc}) or Eq.~(\ref{agc}),
may show wild fluctuations as a function of position or energy.
Indeed, in a ballistic system the DOS is given by a series of closely
spaced $\delta $-function peaks due to the space quantization in the
loop. On a low-spatial resolution, when loops with slightly different
${\cal L}$ contribute to observables, DOS becomes an essentially
smooth function. The summation over the loops in a small interval of
$\delta {\cal L}$ is equivalent to averaging with respect to the phase
of $\varphi = p_{F}{\cal L}$ provided $p_{F}\;\delta {\cal L} > 1$ (cf
\cite{Nag98}).

In the representation Eq.~(\ref{agc}) the integration is easily
performed and the coarse-grained Green's function $\hat{g}_{coarse}$
is simply
\begin{equation}
\hat{g}_{coarse} =
\;\psi_{+}\psi_{+}^{\ddagger}
-
\;\psi_{-}\psi_{-}^{\ddagger}
\; .
\label{cgc}
\end{equation}
For a homogeniuos loop, this expression coincides with that for an
infinite trajectory.

This result is readily generalised to the inhomogeneous loop.  Indeed,
the evolution matrix can be always presented as in Eq.~(\ref{7fc})
because it is the only form compatible with Eq.~(\ref{qfc}). Unlike
the homogeneous case, $\psi_{\pm}$ are $x$-dependent and there is no
simple way to find them. Operationally, $\psi_{\pm}$ are
eigenfunctions of $U_{x}$ and $u$ and $1/ u$ its eigenvalues (the latter are
global $x$-independent values). Without lost of generality, $|u|<1$
and $\psi_{+}$ is the eigenfunction corresponding to the smallest
eigenvalue of $U_{x}$. Then, again one comes to Eq.~(\ref{cgc}). Of
course, the effective wave functions $\psi_{\pm}$ depend on the
coordinate as well as the length the loop ${\cal L}$. The normalization
condition is granted by the orthonormality relation in Eq.~(\ref{4fc}). 

We see that the coarse-grained averaging eliminates ``wrong'' paths in
Eq.~(\ref{fgc}) and the {\em averaged} Green's function can be found
from the standard quasiclassical theory.  However, as we see later,
alternative paths may survive the averaging in more complicated
geometries.

\section{Double loop}\label{2loop}

\begin{figure}[h] 
\includegraphics[height=0.3\textwidth]{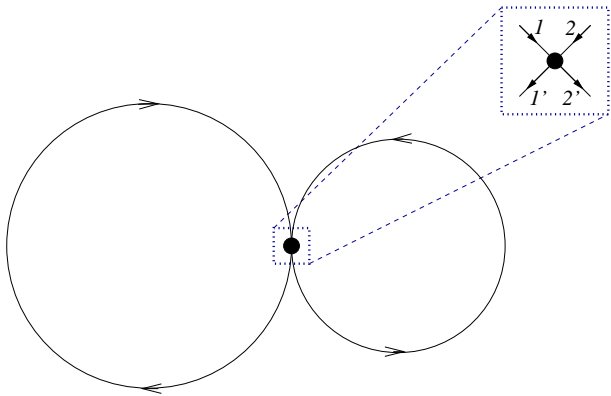}
\caption{Two coupled loops. The arrows show the direction in which
the Fermi phase increases (the  momentum direction). The filled circle
stands for  the knot which ties the loops allowing for the coupling of
the loops.   
The amplitudes of the waves at the points adjacent to the knot,
1, 1' on the left loop and 2, 2' on the right loop, 
are related by the $S$-matrix in accordance with Eq.~(\ref{ggc}).}
\label{2loopp}
\end{figure}

In this section, we generalise the theory to the case of two loops
tied by a knot, 
see Fig. \ref{2loopp}.

As before, we denote the propagator $\hat{g}(x_{1},x_{2})$ with the
understanding that $x_{1}$ and $x_{2}$ may stand for either $x ^{(l)}$
or $x^{(r)}$, the coordinates of the points on the left and right
loops, respectively.  The initial and final points of the propagator
may be on the same loop or belong to the different loops, see
Fig. \ref{2loopp}.  The coordinates $x^{(l,r)}$ are counted from the
knot in the direction of the corresponding arrow. They assume values
$0< x^{(l,r)}< {\cal L}^{(l,r)}$, ${\cal L}^{(l,r)}$ being the
circumference of the left or right loop.  The equations of motion
within each of the loops are given by Eq.~(\ref{vra}).  We couple the
loops with a phenomenological unitary $S$-matrix,
\begin{equation}
S=
\left(
\begin{array}{lr}
 \rho  &  t  \\
 t  &  \rho
\end{array}
\right)
\label{uhc}
\end{equation}
where parameter $t$ describes the inter-loop coupling (``transmission
amplitude'') and $\rho $ is the intra-loop scattering amplitude.
These parameters are taken same for the electron and hole components
of the wave function.  Apart from the the unitarity condition,
\[
|\rho|^{2} + |t|^2 =1 \quad,\quad  \rho t^{*} + \rho^{*} t =0
\;,
\]
$\rho $ and $t$  are free  parameters.
In the limit $t=0$, one has two independent loops, and, if $r=0$, the
two loops are equivalent to a single loop (with the shape like ``8'').

Denote $\psi_{1,1',2,2'}$ the (two-components) wave functions at the
corresponding point 1,1',2 or 2', adjacent to the knot coupling the
loops (see Fig.\ref{2loopp}).  Then, the $S$-matrix expresses the
amplitudes of the outgoing waves, $\psi_{1',2'}$,
\begin{equation}
{\psi_{1'}\choose \psi_{2'}}
=
\left(
\begin{array}{lr}
 \rho  &  t  \\
 t  &  \rho
\end{array}
\right)
{\psi_{1}\choose \psi_{2}}
\label{ggc}
\end{equation}
via the incoming wave amplitudes $\psi_{1,2}$.

The formal solution in Eq.~(\ref{xfc}) for the one-point Green's
functions $\hat{g}_{\pm}(x^{(l,r)})$ remains valid if the loop
evolution operator ${\cal U}_{x^{(l,r)}}$ properly accounts for the
knot.

Consider for definiteness the right loop. The full loop evolution
matrix can be written as the product,
\begin{equation}
{\cal U}_{x^{(r)}} = e^{i p_{F} {\cal L}^{(r)}}
U^{(r)}(x^{(r)},0)\;
{\cal R}^{(r)}\;
U^{(r)}( {\cal L}^{(r)}, x ^{(r)})\; ,
\label{igc}
\end{equation}
where $U^{(r)}(x^{(r)}, x^{(r)\prime})$ is the in-loop evolution
matrix found from Eq.~(\ref{pfc}), and ${\cal R}^{(r)}$ (the transfer
matrix) accounts for the propagation through the knot from the point 2
to the point 2' (see Fig. \ref{2loopp}). The latter can be evaluated
with the help of Eq.~(\ref{ggc}) by the following arguments.

While calculating the one-point Green's function for the right loop,
the waves in the left loop obey homogeniuos equation of motion.  Then,
the amplitudes $\psi_{1}$ (see Fig.\ref{2loopp}) and $\psi_{1'}$ are
related as
\begin{equation}
\psi_{1}= {\cal U}^{(l)} \psi_{1'} 
\label{jgc}
\end{equation}
where
\begin{equation}
 {\cal U}^{(l)}= 
e^{i p_{F} {\cal L}^{(l)}} U^{(l)}
\label{kgc}
\end{equation}
and $U^{(l)}$ is the quasiclassical evolution matrix in the left loop
which advances $x ^{(l)}$ from $x ^{(l)}=0$ to $x^{(l)}= {\cal
L}^{(l)}$:
\[
U^{(l)}= U^{(l)}({\cal L}^{(l)},0)\; .
\]
Combining Eqs.~(\ref{ggc}), and (\ref{jgc}), one excludes amplitudes
$\psi_{1}$ and $\psi_{1'}$, and finds that
\[
\psi_{2'} = {\cal R}^{(r)}\psi_{2}
\]
where
\begin{equation}
{\cal R}^{(r)} =
 {1 - {{\cal U}^{(l)}\over \rho^{*}}\over {1\over \rho} -  {\cal U}^{(l)}}\; .
\label{ngc}
\end{equation}
Eq.~(\ref{vra}) together with Eqs.~(\ref{igc}), and (\ref{ngc}),
allows one to find the Green's function at any point of the loop.

For instance, at point $2'$ and $2$ where $x^{(r)}=0$ and
$x^{(r)}={\cal L}^{(r)}$, respectively,
\begin{eqnarray}
{\cal U}_{2'}     & =  &
\left(1 - {T\over R}\;  
{{\rho} {\cal U}^{(l)}\over 1 - \rho {\cal U} ^{(l)}} \right)
\rho\;{\cal U} ^{(r)}
              \label{rgc}\\   
{\cal U}_{2}     & =  &  
\rho\;{\cal U} ^{(r)}
\left(1 - {T\over R}\;  
{\rho {\cal U}^{(l)}\over 1 - \rho {\cal U} ^{(l)} }\right)
            \label{rgc2}
\end{eqnarray}
where $T = 1- R$.
  
Although not obvious from the beginning, the structure of the
equations allows one to absorb the phase of $\rho $ to the phase
factors $e^{i p_{F}{\cal L}^{(l,r)}}$, so that $\rho $ becomes
effectively a real positive parameter
\[
\rho  \rightarrow  \sqrt{R} \quad,\quad  R = |\rho |^2 \; .
\]

Then Eqs.~(\ref{rgc}), and (\ref{rgc2}) can be presented as
\begin{eqnarray}
{\cal U}_{2}     & =  &
\sqrt{R}\;{\cal U} ^{(r)} - T
\;{\cal U} ^{(r)} 
 {{\cal U}^{(l)}\over 1 - \sqrt{R}\; {\cal U} ^{(l)}} 
\;,
\nonumber 
\\
{\cal U}_{2'}     & =  &
\sqrt{R}\;{\cal U} ^{(r)} - T
 {{\cal U}^{(l)}\over 1 - \sqrt{R} \;{\cal U} ^{(l)}}
\;{\cal U} ^{(r)} 
\;.
\nonumber 
\end{eqnarray}

Inserting these expressions for the full evolution matrices into
Eq.~(\ref{xfca}), one gets after some transformation the Green's
function $\hat{g}_{+}$ at the points $2$ and $2'$:
\begin{widetext} 
\begin{eqnarray}
\hat{g}_{+2}& =  & \left( 1 + T 
{{\cal U}^{(r)}\over 1 - \sqrt{R}\; {\cal U} ^{(r)}}
\;
{{\cal U}^{(l)}\over 1 - \sqrt{R}\; {\cal U} ^{(l)}}
\right)^{-1}
\;\;  
{1\over 1 - \sqrt{R}\; {\cal U} ^{(r)}}    
\rule[-4ex]{0ex}{0ex}
       \label{vgc}\\   
\hat{g}_{+2'}& =  & 
{1\over 1 - \sqrt{R}\; {\cal U} ^{(r)}}\;\;
\left( 1 + T 
{{\cal U}^{(l)}\over 1 - \sqrt{R}\; {\cal U} ^{(l)}}
\;
{{\cal U}^{(r)}\over 1 - \sqrt{R}\; {\cal U} ^{(r)}}
\right)^{-1}
 \label{vgc2}  
\end{eqnarray}
\end{widetext}
These expressions for the Green's function are exact within the
Andreev approximation where the normal metal spectrum is linearised in
the vicinity of the Fermi surface; by a different method, similar
formulae for a sandwich have been previously derived in \cite{Nag98}.
In particular, they allow for the space quantisation due to the normal
reflection (in addition to the Andreev levels grasped by the
conventional quasiclassical theory \cite{Eil68,LarOvc}). The evolution
matrices ${\cal U}^{(r,l)}$ defined by Eq.~(\ref{kgc}), can be
evaluated in the standard quasiclassical scheme, for instance, in the
Riccati equation technique (see Sect.\ref{U}).

\subsection{Checking the normalization condition}\label{checknorm}

Now, we want to demonstrate that unlike the simple loop case, the
normalization condition for the Green's function for a double loop is
violated even on the coarse-grained level.

As before, the coarse grain averaging amounts to the integration of
the Green's functions with respect to the phases factors
\begin{equation}
e^{i\varphi ^{(l,r)}}= e^{i p_{F}{\cal L}^{(l,r)}}
\label{8hc}
\end{equation}
${\cal L}^{(l,r)}$ being the length of the corresponding loop.
In principle, the integration can be performed in terms of the
elliptic integrals as in \cite{Nag98}.  Instead, we use an approximate
procedure which allows one to get better
control 
of the physics
behind the violation of normalization.

Analogously to Eq.~(\ref{7fc}), we present the evolution matrices
entering Eq.~(\ref{vgc}) as
\[
{\cal U}^{(l,r)}= 
e^{i p_{F} {\cal L} ^{(l,r)}}
\left(
u \; \psi_{+}\psi_{+}^{\ddagger}
 +  {1\over u}\;\psi_{-}\psi_{-}^{\ddagger}
 \right)^{(l,r)} \; 
\]
where $u ^{(l,r)}$ are given by Eq.~(\ref{2hc}) for each of the loops.
Algebra is greatly simplified if one uses the matrix representation
where the evolution in the right (or left) loop is diagonal, {\it
i.e.}
\begin{equation}
{\cal U}^{(r)} = 
e^{i p_{F} {\cal L} ^{(r)}} \left(
\begin{array}{lr}
u^{(r)}   & 0   \\
 0  & {1\over u^{(r)}}
\end{array}
\right)\; .
\label{9gc}
\end{equation}
This can be always achieved by the transformation of all the matrices,
\begin{equation}
\hat{X} \rightarrow {\cal O} \hat{X} {\cal O}^{-1}
\label{yhc}
\end{equation}
with the matrix ${\cal O}$, which diagonalizes ${\cal U}^{(r)}$.  In
this representation, the right loop is effectively a normal metal.

Up to a normalisation factor, the eigenfunction $\psi_{\pm}^{(l)}$ of
${\cal U}^{(l)}$ can be written as: $$\psi_{+}^{(l)}= {1 \choose
A_{+}} \quad,\quad \psi_{-}^{(l)}= {A_{-} \choose 1}.$$ In {\em this}
representation, $A_{\pm}$ has a simple physical meaning: $A_{+}$
($A_{-}$) are amplitudes of the Andreev reflection of the electron
(hole) excitation for the ideal case $t=1$, $\rho =0$ (a totally
transparent interface).

The left layer evolution matrix reads
\begin{equation}
{\cal U}^{(l)} = 
e^{i p_{F} {\cal L} ^{(l)}}
\left(
{1\over 2}\left( 1 + g_{0}^{(l)}\right) u^{(l)}
+
{1\over 2}\left( 1 - g_{0}^{(l)}\right) \frac{1}{u^{(l)}}
 \right)
\label{ahc}
\end{equation}
where
\begin{equation}
g_{0}^{(l)}
 = {1\over 1- A_{+}A_{-}}
\left(
\begin{array}{lr}
1 + A_{+}A_{-}  & -2 A_{-}\\
2 A_{+}& -1 - A_{+}A_{-}
\end{array}
\right) 
\;.
\label{bhc}
\end{equation}

We plug these relations into Eq.~(\ref{vgc}) and perform an expansion
in series of $u ^{(l,r)}$. From Eqs.~(\ref{9gc}), and (\ref{bhc}), one
sees that each power of $u ^{(l,r)}$ brings with it $e^{+ i \varphi
^{(l,r)}}$ (electron) or $e^{- i \varphi ^{(l,r)}}$
(hole). Accordingly, the expansion for a Green's function $G$ has the
following structure
\begin{equation}
G = \sum\limits_{k,n = 0}^{\infty } 
(u^{(l)})^{k}(u^{(r)})^{n}
\sum\limits_{p= -k}^{k} \sum\limits_{q=-n}^{n}
e^{ip \varphi^{(l)}}
e^{iq \varphi^{(r)}}
G_{k,n}^{p,q}
\label{7hc}
\end{equation}
Each term which can be associated with the corresponding process where
the particle starts moving from an initial point and returns back
having moved around the loops. {\em En route}, it undergoes
electron-hole conversion as a result of the Andreev reflection on the
knot.  The Fermi momentum phase increases when the particle moves in
the direction of the arrows in Fig. \ref{2loopp} (as the electron) and
decreases otherwise (for the hole).  The total acquired phase is
controlled by the history of the electron-hole conversion.  The
meaning of the integers $k,n,p,$ and $q$ in in Eq.~(\ref{7hc}) is the
following. For instance, $k$ shows the total length of the path $k
{\cal L}^{(l)}$ in the left loop. If the particle have circled the
loop $k_{e}$ times as electron and $k_{h}$ times as hole ($k= k_{e}+
k_{h}$), the acquired Fermi momentum phase corresponds to $p$ in
Eq.~(\ref{7hc}) equal to $p= k_{e}- k_{h}$.  Parameters $n$ and $q$
have analogous meaning for the right loop.

On the coarse-grain level of description, one performs averaging
within an interval of the loop length $\delta {\cal L}\ll {\cal L}$
which is wide enough to average the phase factors in Eq.~(\ref{8hc})
to zero.  Independent averaging with respect to length of the left and
right loops, amounts to ignoring any term in Eq.~(\ref{7hc}) other
than $p=q=0$.  For certain geometries, the phase factors $e^{i
\varphi^{(l)}}$ and $e^{i \varphi^{(r)}}$ cannot be considered
independent. This occur if ${\cal L}^{(l)}= {\cal L}^{(r)}$, or more
generally, the lengths of the loops are commensurable, $N {\cal
L}^{(l)} = M {\cal L}^{(r)}$ , $N,M=1,2\ldots$.  In this case, the
terms $p =\pm N, q= \mp M $ (and their multiples) also contribute to
the coarse grain average.

For definiteness, we consider the Green's function, $\hat{g}_{2}$,
corresponding to the point 2 in the immediate vicinity of the knot
(see Fig. \ref{2loopp}). The loops are assumed to be incommensurable.
The result of the calculation of the averaged Green's function
$\langle \hat{g}_{2}\rangle $ can be presented as follows:
\begin{equation}
\langle \hat{g}_{2}\rangle = 
\hat{G}_{qc} + \hat{G}_{nqc}
\label{9hc}
\end{equation}
where $\hat{G}_{qc}$ is of the quasiclassical Maki-Schopohl form,
\[
\hat{G}_{qc} = 
\frac{1}{1- \alpha \beta }
\left(
\begin{array}{lr}
1 + \alpha \beta    & -2 \alpha    \\
 2 \beta   & -(1- \alpha \beta )
\end{array}
\right)
\]  
with
\begin{eqnarray}
\alpha      & =  &
  A_{-}\,\gamma\,T\, {{u^{(r)}}^2}\left( 1 - 
\gamma^2 \mu \nu  T {{u^{(l)}}^2}  
       \right)
+ \ldots
\nonumber 
\\   
\beta      & =  &  
A_{+}\,\gamma\,T\,\left( 
     \left( 1 - \gamma^2
\mu 
    \nu 
         T\,{{u^{(l)}}^2} 
   \right)
      \left( 1 + \gamma^2 
\mu^2\,
         R\,{{u^{(r)}}^2} 
      \right)   \right. \nonumber \\
&& +
\left. 
T^2\gamma^4{u^{(l)}}^2\,{u^{(r)}}^2
\mu^2
      \left(  
\nu^2 
+ 4 R\,A_{-}\,A_{+} \right) \,
       \right) 
+ \ldots
            \label{aic2}   
\end{eqnarray}
where
\[
\mu = 1 - A_{+}A_{-}
\; ,\;  
\nu = 1 + R A_{+}A_{-}
\; ,\;  
\gamma  = \frac{1}{1 - R A_{+}A_{-}} 
\]
these combinations of the Andreev amplitudes are well-known in the
theory of an imperfect NS-interface \cite{She80b}.  The second term in
the r.h.s. of Eq.~(\ref{9hc}), $\hat{G}_{nqc}$, reads
\begin{eqnarray}
\hat{G}_{nqc}& =&
4
 \gamma^{4}
\mu^2
   R T^2
u^{(r)2}u^{(l)2} \nonumber \\
&\times&
\left(
\begin{array}{cc}
A_{+}A_{-}   & 0   \nonumber\\
A_{+} \gamma 
\left( 2T\,A_{+}\,A_{-}  - \mu \nu  \right)\hspace*{1ex} 
  &
    - A_{+}\,A_{-}
       \end{array}
\right) + \ldots
\; .
\label{cic}
\end{eqnarray}

Squaring the coarse-grained Green's function, one gets
\begin{equation}
\langle g_{2}\rangle^{2} - \hat{1} =
  8\,A_{+}\,A_{-}\,
   \mu ^2
\,{{\gamma}^4}\,
   R\,{T^2}\,{{u^{(l)}}^2}\,{{u^{(r)}}^2}  \; \hat{1} + \ldots
\label{4gc}
\end{equation}
These expressions are valid up to $u^{(r)2}$ and $u^{(l)2}$, their
product included.

One sees that the normalization condition is {\em not} restored by the
coarse-grain averaging as in the case of a single loop. From the
structure of the r.h.s. of Eq.~(\ref{4gc}), one may infer that the
process responsible for the violation of the normalization condition
is the one where the particle follows an $\infty $-like path twice
along the loops, each in the opposite direction; such a path is
$(u^{(l)}u^{(r)})^2$ long and the change of the direction of motion
requires two sequential Andreev reflections, $A_{+}$ and $A_{-}$.  In
the limits $R \rightarrow 0$ or $T \rightarrow 0$, one restores the
single loop situation and the normalization condition.  Note also, the
normalization is not violated if $A_{\pm}=0$, {\it i.e.}  the
electron-hole conversion is absent (in the normal metal case or if the
loops are made of the same material).

\section{Sandwich}\label{sandwich}

Now we are in position to apply the theory of previous sections to a
physical system. We consider a two-layer system shown in
Fig.\ref{2layer} formed by two superconducting layer, left and
right. The properties of the layers are specified by their thicknesses
$a ^{(l,r)}$ and the complex order parameter $\Delta ^{(l,r)}$ (they
are considered as input, and the self-consistency condition is not
discussed).  We assume that the order parameter depends only on the
coordinate $x$ perpendicular to the layer plane.

The Gor'kov equation for the matrix Green's function ${\Bbb
G}_{\varepsilon }(\bm{r}_{1}, \bm{r}_{2})$  reads
\[
(\varepsilon - {\mathbb H} - 
{\mathbb V}
) {\mathbb G}_{\varepsilon}=  
\openone
\]
where ${\mathbb H}$ is defined in Eq.~(\ref{wec}) and ${\mathbb V}$ is
the potential describing the the interface.  As before, we assume the
parabolic spectrum, {\it i.e.}
\[
\hat{\xi } = {\hat{\bm{p}}^{2}\over 2m} - {p_{F}^{2}\over 2m}\;.
\]

Due to the in-plane translational symmetry, the Green's function can
be presented as
\[
{\mathbb G}_{\varepsilon}(\bm{r},\bm{r}') =
\sum\limits_{\bm{p}_{||}} 
{\mathbb G}_{\varepsilon }(x,x';
\bm{p}_{||})
e^{i \bm{p}_{||\cdot}(\bm{r}-\bm{r}')_{||}} \; .
\]
To simplify notations, we use below ${\mathbb G}(x,x')$ as a shorthand
for ${\mathbb G}_{\varepsilon }(x,x'; \bm{p}_{||})$.

The matrix $2\times 2$ Green's function $ {\mathbb G}(x,x')$ is found
from
\begin{equation}
(\varepsilon - {\mathbb H}_{x})
{\mathbb G}(x,x') =  \hat{1} \delta (x-x') 
\label{4yb}
\end{equation}
where 
\[
{\mathbb H}_{x}=
\left(
\begin{array}{cc}
\varepsilon -\xi_{x} & -\Delta \\
-\Delta^{*} & \varepsilon + \xi_{x}
\end{array}
\right)
+ {\mathbb V}(x)
\]
 and 
\[
\xi_{x}= {1\over 2m} \left({1\over i} {d \over{dx}} \right)^2 -
{p_{Fx}^{2}\over 2m}
\quad,\quad 
{p_{Fx}^{2}\over 2m}= {p_{F}^{2}\over 2m} \cos^2 \theta 
\]
where
\[
\cos^2 \theta = \left(1 - \frac{{\bm{p}}_{||}^{2}}{p_{F}^2} \right)
\]

The boundaries at $x = - a ^{(l)}, a ^{(r)}$ are assumed impenetrable
so that
\begin{equation}
{\mathbb G}(x_{1}, {x_{2}=- a ^{(l)} \text{ or } a ^{(r)}})=
{\mathbb G}({x_{1}=- a ^{(l)} \text{ or } a ^{(r)}}, x_{2})= 0
\label{gic}
\end{equation}

The Green's function $\hat{G}(x,x')$ is continuous at $x= x'$
\[
{\mathbb G}(x + 0, x)=
{\mathbb G}(x - 0, x) \quad ,
\]
 whereas its derivatives suffer a jump generated by the delta-function
in the r.h.s. of Eq.(\ref{4yb}):
\[
{i \hbar \over 2m}\left[p\, \hat{G}\hat{\tau}_{z} \right]_{x =
x^{'-0}}^{x = x^{'+0}} = 1
\]

If one chooses a particular model for the interface between the
layers, e.g. a $\delta $-function barrier, one can solve the Gor'kov
equation for the retarded Green's function exactly.  The analysis of
the full solution shows that the Andreev approximation, greatly
simplifying the algebra, faithfully preserve all the physics we are
going to discuss. Therefore, we restrict ourself to the Andreev
approximation as in Sect. \ref{two-point} and \ref{2loop}.

As in Section \ref{two-point}, we split the Green's function into two
pieces Eq.~(\ref{7ec}), $\mathbb G^{+}$ and $\mathbb G^{-}$, which
describe propagation of the particle injected in the system with the
momentum $+ p_{Fx}$ and $- p_{Fx}$, respectively.  Consider for for
definiteness the Green's function $\mathbb G^{+}$ in the right layer.
Since the initial direction of the momentum may be changed by the
reflection on the outer walls and the interface, instead of
Eq.~(\ref{9ec}) one has
\begin{eqnarray}
i v_{F} \;{\mathbb G}^{+}(x_{1},x_{2}) \tau_{z} & =&
g^{++}( x_{1},x_{2}) \; e^{i p_{Fx}(x_{1}-x_{2})}
\nonumber \\
&-&
g^{+-}( x_{1},x_{2}) \; e^{- i p_{Fx}((x_{1} +x_{2} - 2 a^{(r)})}
\;;
\label{dic}
\end{eqnarray}
the function $g^{++}$ suffers the jump Eq.~(\ref{bfc}) at the source
point $x_{1}= x_{2}$ whereas the secondary wave $g^{+-}$ is
continuous.

\begin{figure}[h] 
\includegraphics[height=0.12\textwidth]{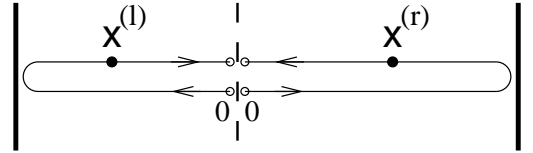}
\caption{
The trajectory coordinate. In each of the layers the trajectory
coordinate ${\sf x}^{(l)}$ and ${\sf x}^{(r)}$ are counted from the
interface along the path of a particle which moves from the interface
and is reflected on the surface boundary.} 
\label{lulu}
\end{figure}

From the boundary condition Eq.~(\ref{gic}),
\begin{equation}
g^{++}( x_{1}= a ^{(r)},x_{2})= 
g^{+-}( x_{1}= a ^{(r)},x_{2}) \; .
\label{hic}
\end{equation}
Because of the reflections, the particle moves 
along the
trajectories shown in Fig. \ref{lulu}. 

It is convenient to introduce the trajectory coordinates ${\sf
x}^{(l,r)}$, which assume values in the region $0< {\sf x}^{(l,r)}<2a
^{(l,r)}$, as
\begin{eqnarray}
{\sf x}^{r}&=& \left\{
\begin{array}{rcr}
x   &\;,\;&  0<{\sf x}^{r} < a ^{(r)}\;;\nonumber\\
2 a ^{(r)}-x   &\;,\;& a ^{(r)}< {\sf x} <2 a ^{(r)}       \;.
\end{array}
\right.
\nonumber \\
{\sf x}^{l}&=& \left\{
\begin{array}{rcr}
-x   &\;,\;&  0<{\sf x}^{l} < a ^{(l)}\;;\nonumber\\
2 a ^{(l)}+ x   &\;,\;& a ^{(l)}< {\sf x} <2 a ^{(l)}       \;.
\end{array}
\right.
\nonumber 
\end{eqnarray}
Since the trajectory coordinate space is twice as large as the
$x-$space, the two functions of $x$ in Eq.~(\ref{dic}) may be packed
into a single function of the trajectory coordinates ${\sf x}_{1}$ and
${\sf x}_{2}$, that is
\begin{widetext}
\[
i v_{F} \;{\mathbb G}({\sf x}_{1},{\sf x}_{2}) \tau_{z}
=
\left\{
\begin{array}{lcr}
g^{++}( x_{1},x_{2}) \; e^{i p_{F}(x_{1}-x_{2})}|_{x_{1,2}= {\sf
  x}_{1,2}}&\;\;,\;\;& 
0<{\sf x}_{1}< a ^{(r)}\;;\\
g^{+-}( x_{1},x_{2}) \; e^{- i p_{F}((x_{1}-x_{2} - 2
  a^{(r)})}|_{x_{1}=2 a ^{(r)} - {\sf x}_{1}, x_{2}= {\sf x_{2}}}
   &\;\;,\;\;&     
a ^{(r)}<{\sf x}_{1}< 2a ^{(r)}\;;\\
\end{array}
\right.
\]
\end{widetext}
for ${\sf x}_{2}< a ^{(r)}$.  Similar relations, express ${\mathbb
G}^{-}( x_{1},x_{2})$ via ${\mathbb G}( {\sf x}_{1}, {\sf x}_{2})$ for
the value of ${\sf x}_{2}$ in the region $a ^{(r)}< {\sf x}_{2}< 2 a
^{(r)}$.  The trajectory in the left loop is introduced in the same
manner.

Finally, one comes to the representation identical to that introduced
for the loop Green's function by Eq.~(\ref{kfc}), {\it i.e.}
\[
i v_{F} \;{\mathbb G}({\sf x}_{1},{\sf x}_{2}) \tau_{z}= 
g( {\sf x}_{1},{\sf x}_{2}) \; e^{ i p_{Fx}({\sf x}_{1}-{\sf x}_{2})}\;.
\]
By virtue of Eq.~(\ref{hic}), the quasiclassical envelope $g( {\sf
x}_{1},{\sf x}_{2})$ is continuous at the point of the loop
corresponding to the outer walls. The envelope satisfies the
quasiclassical equations Eq.~(\ref{vra}) with the understanding
$x_{1,2}$ is the trajectory coordinate ${\sf x}_{1,2}$.

The coupling of the left and right layers can be described
phenomenologically with the help of the S-matrix Eq.~(\ref{uhc}).  The
simple structure of the S-matrix in Eq.~(\ref{uhc}) corresponds to the
assumption the interface is mirror symmetric \cite{delta-pot}.  In
this context, $\rho $ is the amplitude of the reflection by the
interface and $t$ is the transmission amplitude.  Then, the theory of
the loop Green's function presented in Sections \ref{loop} and
\ref{2loop} can be directly applied to the case of a layer or a
sandwich.  Within then Andreev approximation, this gives the solution
which is ``exact'' in the sense that it allows for the space
quantization (in \cite{Nag98}, this Green's function is called {\em
quasiclassical-like}).

\begin{figure}[h] 
\includegraphics[height=0.25\textwidth]{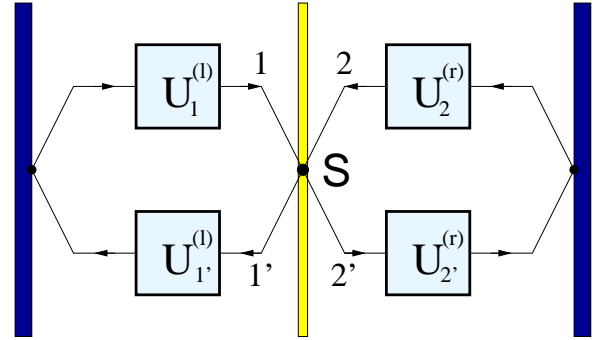}
\caption{Waves with definite in-plane momentum in a  sandwich. The
wave is totally reflected by the outer waves. The interface is
partially transparent. To show the two-component electron-hole
character of the wave, we use the double lines. The evolution
operators $U$'s mix electron-hole components whereas the interface
scattering is described by an electron-hole scalar $S$-matrix.}
\label{sand-circuit} 
\end{figure}

Schematically, the loops are shown in Fig. \ref{sand-circuit}: Apart
from the Fermi momenta phase factor, the propagation of the waves
moving towards the interface is described by the evolution operator
$U_{1}^{(l)}$ and $U_{2}^{(r)}$ in the left and right layer,
respectively. The evolution of the outgoing waves is controlled by
$U_{1'}^{(l)}$ and $U_{2'}^{(r)}$. The in- and out-waves are mixed by
the S-matrix, an electron-hole scalar in the Andreev approximation.

The Green's function can be then found from Eq.~(\ref{vgc}) and
\ref{vgc2} where 
\begin{eqnarray}
{\cal U}^{(l,r)}& =&  e^{(2 p_{Fx}a^{(l,r)}i+ \pi i) }\; U ^{(l,r)}
\nonumber \\
 U ^{(l)}= U_{1}^{(l)}U_{1'}^{(l)}
&,& U ^{(r)}= U_{2}^{(r)}U_{2'}^{(r)} \; .
\label{ybc}
\end{eqnarray}
The evolution matrix depends on the direction of propagation in an
anisotropic superconductor and for this reason $U_{2}^{(r)}$ and
$U_{2'}^{(r)}$ may differ from each other.  The additional $\pi
$-phase shift is due to the reflection on the wall (the minus sign in
the r.h.s.  of Eq.~(\ref{dic})).  The matrices $U ^{(l,r)}$ can be
found solving usual quasiclassical equations (see Section \ref{U}).

In the simplest case of isotropic homogeneous superconductors, the
evolution matrices are simply
\begin{eqnarray}
{\cal U}^{(l,r)}& =& 
e^{i\pi + 2i p_{F}a^{(l,r)}/ \cos \theta   }
\left(
{1\over 2}\left( 1 + g_{\text{bulk}}^{(l,r)}\right) u^{(l,r)}
\right.
\nonumber \\
&+&
\left.
{1\over 2}\left( 1 - g_{\text{bulk}}^{(l,r)}\right) \frac{1}{u^{(l,r)}}
 \right)
\nonumber 
\end{eqnarray}
where $g_{\text{bulk}}$ stands for the bulk Green's function and 
\[
u ^{(l,r)} =  e^{{2i \xi^{(l,r)} a ^{(l,r)}\over v_{F} \cos \theta  } }
\; ,
\]
$\xi $ is defined by Eq.~(\ref{9fc}) (in the ballistic case, $\xi =
\sqrt{\varepsilon^2 - | \Delta |^{2}}$).

The transformation to the convenient representation Eq.~(\ref{yhc})
where one of the layers (the right one) becomes effectively the normal
metal can be performed with the matrix \cite{SheOza00}
\[
{\cal O} = \left(
\begin{array}{lr}
1   & a_{\text{bulk}} ^{(r)}\\
b_{\text{bulk}}^{(r)}& 1
\end{array}
\right)
\]
where $a_{\text{bulk}} ^{(r)}$ and $a_{\text{bulk}} ^{(r)}$ are the
parameters of the Maki-Schopohl parametrisation of
$g_{\text{bulk}}^{(r)}$.

It follows from the derivation in Sections \ref{loop} and \ref{2loop},
that the exact Green's function, e.g. at the point 2 (see
Fig. \ref{sand-circuit}), $g_{2}$ can be found using Eq.~(\ref{xfcc})
as Eq.~(\ref{xfcc}),
\[
g_{2} = 
 {{1 + {\cal R}^{(r)} {\cal U}^{(r)}}\over 
1- {\cal R}^{(r)} {\cal U}^{(r)}}             
\quad,\quad 
\]
where in accordance with Eq.~(\ref{ngc}) ${\cal R}^{(r)}$ can be
expressed via the reflection amplitude $\rho $ and the evolution
matrix ${\cal U}^{(r)}$.

As already mentioned, the exact propagator given by Eq.~(\ref{vgc})
must not be compared with the quasiclassical Green's function because
they are different physical objects. Indeed, the Green's function in
the Gor'kov equation Eq.~(\ref{4yb}) describes a one-dimensional
propagation of a plane wave with definite parallel momentum
$\bm{p}_{||}$ and, therefore, of infinite extension in the direction
parallel to the sandwich plane.  Being infinite, the multiply
reflected waves inevitably overlap and their interference leads to the
bound state formation (which would be reveal itself as Fabry-P\'erot
type resonances if the sandwich was connected to the outside world).
On the other hand, a beam-like wave packet built of the plane waves
with close values of $\bm{p}_{||}$, which behaves like a classical
particle in a billiard, does not overlap with itself and the
resonances are absent.  The wave packet construction, which produces
the trajectory picture shown in Fig. \ref{coarse}, is equivalent to
the coarse-grained averaging discussed before. The coarse-grained
Green's function for the sandwich is given by Eq.~(\ref{9hc}) at the
point 2 chosen as a representative point.  In what follows we want to
show that in an {\em inhomogenious superconductors} the interference
effects survive the coarse grain averaging.  For this, we find the
quasiclassical Green's function and compare it with exact
coarse-grained solution Eq.~(\ref{9hc}).

\subsection{Quasiclassical Green's function }\label{quasi}

In the framework of the quasiclassical theory, an iteration procedure
which allows one to find the Green's function of a sandwich, has been
suggested in \cite{SheOza00}. Here, we use a modification of the
method which is more convenient for the expansion with respect to the
path length.

We use the Maki-Schopohl parametrisation of the Green's function
\cite{SchMak95},
\[
\hat{g} = {1\over 1- ab} \left(
\begin{array}{lr}
1 + ab   & - 2 a    \\
 2 b  & -1 - ab
\end{array}
\right)\; ,
\]
and the boundary conditions for the Andreev amplitudes $a$ and $b$
\cite{Esc00,SheOza00,OzaShe01} at the interface.

\begin{figure}[h] 
\includegraphics[height=0.3\textwidth]{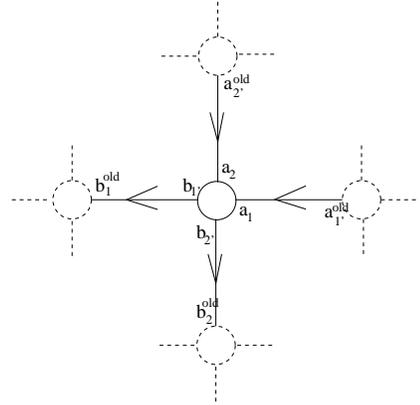}
\caption{Scattering on the interface. The incoming channel are 1 and
2, and the outgoing ones are 1' and 2'. The numbering  is same as in 
Fig. \ref{sand-circuit}. The arrows show the direction of the
momentum.  The input for the interface boundary
conditions is the set of  Andreev amplitudes $a_{1,2}, b_{1',2'}$. 
These parameters can be found from the output of the scattering event
on the previous collision $a_{1',2'}^{\text{old}}$, $b_{1,2}^{\text{old}}$. 
}
\label{iter}
\end{figure}

Consider a collision with the interface, see Fig.\ref{iter}.  In
accordance with \cite{SheOza00,OzaShe01}, the boundary condition can
be conveniently expressed via the determinant
\[
{\cal D}= \det ||1 - S \hat{a} S^{\dagger}\hat{b}||
\]
where $S$ is the interface S-matrix Eq.~(\ref{uhc}), and $\hat{a}=
\text{diag}(a_{1}, a_{2}) $ and $\hat{b}= \text{diag}(b_{1'}, b_{2'})$
are diagonal matrices bulit from the ``input'' Andreev amplitudes at
the trajectory points immediately ajacent to the interface
(Fig. \ref{sand-circuit}).  The determinant reads
\[
{\cal D}= R (1 - a_{1}b_{1'})(1- a_{2}b_{2'})
+
T(1 - a_{1}b_{2'})(1- a_{2}b_{1'})
\;,
\]
where $R$ and $T$ are the reflection and transmission coefficients.
The ``output'' amplitudes $a_{1',2'}$ and $b_{1,2}$ can be found by
the following formulae:
\begin{equation}
  a_{n'} =   - {{\cal D}_{1}^{(n')}\over {\cal D}_{0}^{(n')}} 
\quad,\quad 
  b_{l} =   -  {{\cal D}_{1}^{(l)}\over {\cal D}_{0}^{(l)}} 
\quad,\quad n,l = 1,2
\; 
\label{u3b}   
\end{equation}
where we denote 
\begin{eqnarray}
{\cal D}_{0}^{(l)}&=& {\cal D}|_{a_{l}=0} \;, \nonumber\\  
{\cal D}_{1}^{(l)}&=& {\partial\over{\partial a_{l}}}{\cal D}\; ,
\nonumber\\
{\cal D}_{0}^{(n')}&=& {\cal D}|_{b_{n'}=0} \; ,\nonumber\\
{\cal D}_{1}^{(n')}&=& {\partial\over{\partial b_{n'}}}{\cal D}\;\; .
\nonumber 
\end{eqnarray}

In the iteration procedure illustrated by Fig. \ref{iter}, the updated input amplitudes, $a_{1,2}$
and $b_{1',2'}$, are found from the output amplitudes of the previous
step, $a_{1',2'}^{\text{old}}$ and $b_{1,2}^{\text{old}}$, from the
following relations:
\[
{a_{1,2}\choose 1} \propto {\cal U}^{(l,r)}{a_{1',2'}^{\text{old}}\choose 1}
\quad,\quad 
{1\choose b_{1',2'}} \propto {\cal U}^{(l,r)}{1\choose b_{1,2}^{\text{old}}}
\; ,
\]
where the ``bond'' evolution operators ${\cal U}^{(l,r)}$ are those in
Eq.~(\ref{ybc}). These relation are understood in the following sense:
For instance, to update $a_{1}$ from the value $a_{1'}^{\text{old}}$,
one calculates the column ${\cal U}^{(l)} {a_{1'}^{\text{old}} \choose
1}$, with the evolution matrix Eq.~(\ref{ahc}), and the advanced value
$a_{1}$ is found as the ratio of the upper and lower components of the
column.

Given the updated input, the new output values are found from
Eq.~(\ref{u3b}). The initial input is the bulk value of the
parameters. In the following, we use the representation diagonalizing
the bulk Green's function in the right layer (as in Section
\ref{checknorm}). Then the zero step value are $b_{1}= A_{+}$,
$a_{1'}= A_{-}$, $b_{2}=a_{2'}=0$, and the evolution operators are
those in Eqs.~(\ref{9gc}), and (\ref{ahc}).

Unlike the iteration procedure suggested in \cite{SheOza00}, on any
iteration steps the parameters $a$ and $b$ have direct physical
meaning -- they give the Green's function at the central knot of the
tree with $4^{n}+1$ knots, $n=0,1,\ldots$ being the number of the
iteration steps.

The result of the calculation is the following.  In the lowest order
with respect to the path length, we recover the exact coarse-grained
Green's function discussed in Sect. \ref{sandwich} and derived in
Sect. \ref{checknorm}. However, the quasiclassical theory reproduces
{\em incorrectly} terms which are proportional to the product
$u^{(r)2} u^{(l)2}$, and, of course, higher order terms.
Specifically, the quasiclassical theory faithfully reproduces the
first terms, $\hat{G}_{qc}$, in the exact expression Eq.~(\ref{9hc})
but it misses the second one, that is $\hat{G}_{nqc}$.  We see that
$\hat{G}_{nqc}$ in Eq.~(\ref{cic}) gives the correction to the
quasiclassical approximation in the lowest approximation with respect
to the path length.

\section{Discussion and conclusions}\label{disc}

In the previous section, we have shown explicitly that the
quasiclassical theory gives, strictly speaking, incorrect results when
applied to a sandwich with a specular reflecting interface. By this,
we confirm some previous observations \cite{Nag98}.  We understand
this result as due to differences in the actual topology of the
trajectories and those assumed in the quasiclassical theory.  Our
arguments are as follows

\begin{figure}[htbp]
\centerline{ 
\raisebox{-3ex}{\rlap{\hspace*{.06\textwidth}(a)}} 
\includegraphics[height=0.33\textwidth]{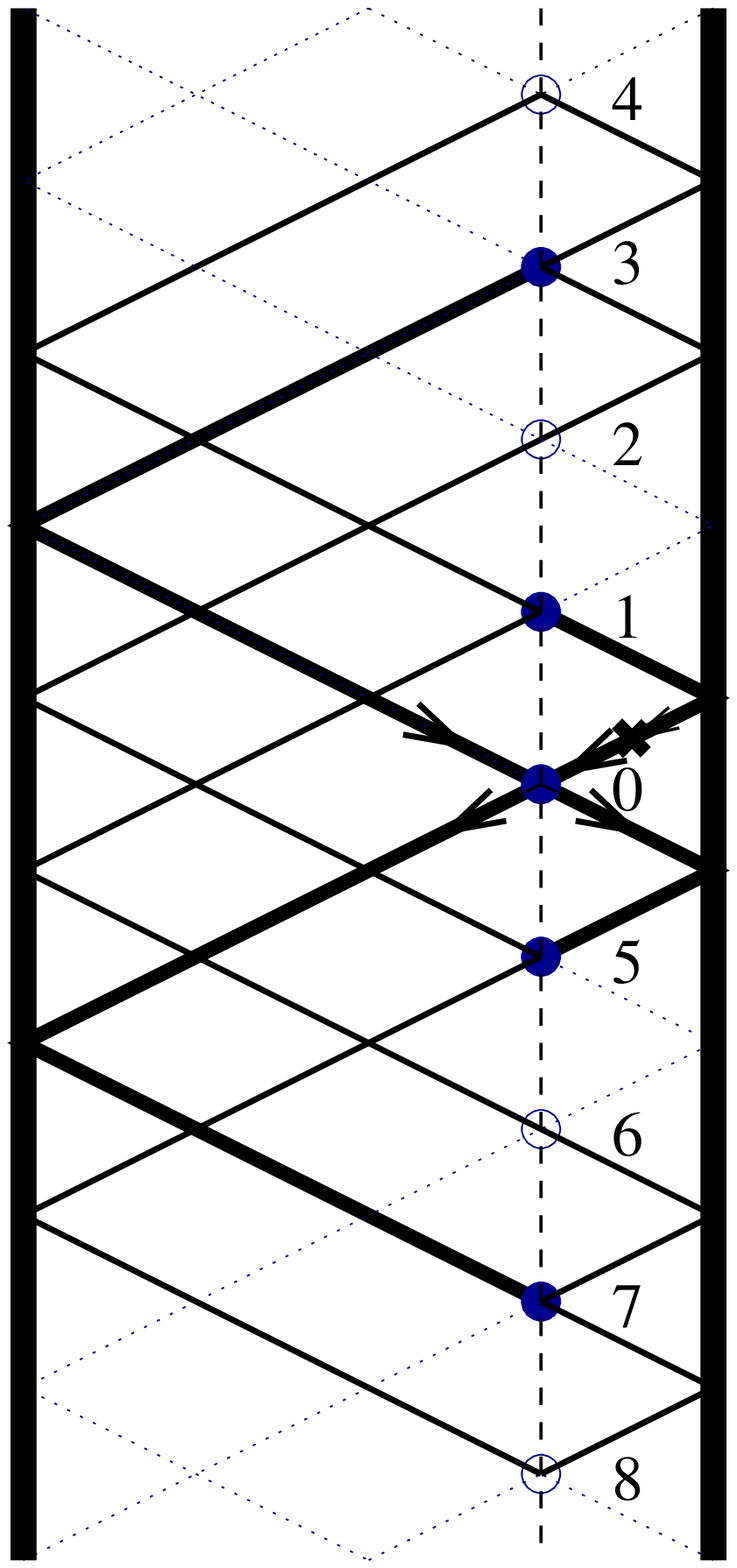}
\hspace*{.025\textwidth}
\raisebox{-3ex}{\rlap{\hspace*{.12\textwidth}(b)}}
\includegraphics[height=0.28\textwidth]{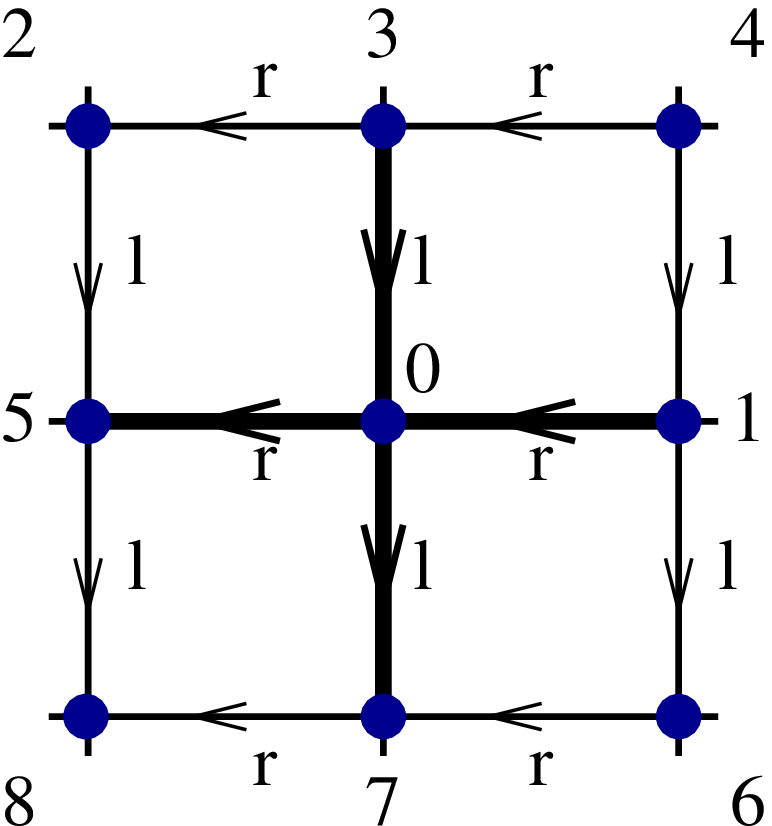}
}
\caption{Typical trajectory in a sandwich. (a) The cross shows the
initial position of the particle near the knot marked by ``0''. Due to
the possibility of the electron-hole conversion, the particle may
change its branch and therefore the direction of motion when hitting
the interface (dashed line). The connectivity of the trajectory is
shown schematically in (b). The knots are numbered as in (a), letters
$r$ and $l$ label the pieces of the trajectory in the right or left
layer, respectively. This picture refers to the general case when the
thickness of the left and right layers are
incommensurable. Fig. \ref{coarse}(a) shows the case of a symmetric
sandwich. }  \label{traj} \end{figure}

In Fig.\ref{traj}, we show fragment of the trajectory space which is
large enough to include all the paths contributing to the approximate
Green's function in Eq.~(\ref{9hc}): the particle starts in the
immediate vicinity of the knot ``0'' and after travelling around
returns back. In the normal metal case, the particle always follows
the arrows being the electron, or always moves against the arrows if
it is the hole, and there are no return paths. Correspondingly,
coarse-grained properties are the same as in the bulk (as it is
well-known from the the theory of normal metal films).  In a
superconductor, the electron-hole conversion is possible due to the
Andreev reflection process when the particle hits the interface
separating the layers with different order parameters.  Then, the
particle may move along the arrows as well as in the opposite
direction.  For instance (see Fig. \ref{traj}(b)), the particle may
move from ``0'' to ``5'' as electron, hit the ``5'' and being
reflected as a hole move back to ``0'' (this give contribution to
$u^{(r)2}$-terms in Eq.~(\ref{aic2})).

\begin{figure}[htbp]
\includegraphics[height=0.3\textwidth]{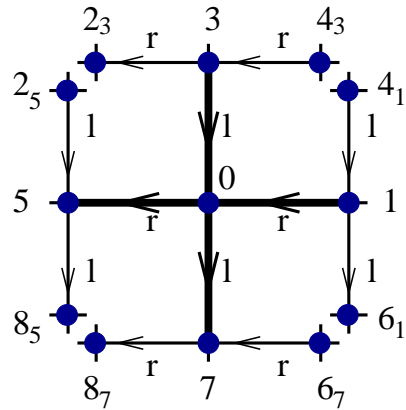}
\caption{Tree like trajectory in the quasiclassical theory. Numbering
of the notes in the same as in Fig. \ref{traj}. The subscript  shows the
knot from which the given knot can be reached.}
\label{trajQuasi}
\end{figure}

The connectivity in the quasiclassical theory is shown in
Fig. \ref{trajQuasi}.  The trajectory is tree-like\cite{SheOza00}
where sequential collisions with the interface are completely
uncorrelated as if the surfaces are rough: For instance, the knot
$2_{3}$ which can be reached from the knot $3$, is regarded as
different from the know $2_{5}$ to which the particle may come from
the knot $5$.

It is clear from Figs. \ref{traj} and \ref{trajQuasi} that the whole
class of the paths allowed in the exact theory is not presented in the
quasiclassical approach. These are loop-like paths, for instance,
''0'' `$\rightarrow$ ``5'' $\rightarrow$ ``2'' $\rightarrow$ ``3''
$\rightarrow $ ``0''. This path includes two Andreev reflections -- at
the knots ``5'' and ``3'', and its length corresponds to $u^{(l)2}
u^{(r)2}$. The Fermi momentum phase factors cancel exactly (in a
translationally invariant system), and therefore, these paths are
immune to the coarse-grain averaging.  Summing over all the loop-like
paths of this length, one is able to account for the r.h.s.  of
Eq.~(\ref{cic}). In the geometry under consideration, the loops are
not rare events (as in the weak localization theory), and, nominally,
their contribution, {\it i.e.} the non-quasiclassical correction, is of
order of 1.  

This analysis, which is the main result of the paper, gives the
physical explanation of the invalidity of the quasiclassical theory in
many layer systems first noticed in Ref. \cite{Nag98} and analyzed
numerically in \cite{OzaShe01}.  The terms missing in the
quasiclassical theory can be definitely attributed to the loops formed
by the interfering paths.  Similar conclusion has been made for the
case of open geometry calculating the Josephson current through a
layered system \cite{OzaSheTob01}.  The interference survives
coarse-grain averaging because the lengths of the two intefering paths
remain equal for any initial condition.  In turn, the loops, as closed
orbits in the billiard theory, are possible only in very special --
integrable -- geometries.  Here it is important that the sandwich has
a perfect in-plane homogeneity and the reflection is specular. Any
imperfection on the scale larger than the Fermi wave length (but,
perhaps, negligible small compared with the coherence length),
transforms the highly connective trajectory in Fig. \ref{traj}(a) into
the tree in Fig. \ref{trajQuasi}. One may say that the quasiclassical
theory is stable relative to small variations of geometry and is
applicable in most of the cases. However, there exist physically
realizable situations where the conventional quasiclassical theory is
inapplicable due to the contribution of the intefering paths.  In this
case, one should use the modification of the quasiclassical technique
suggested in the present paper.

Finally, we want to mention that the physics due to the loops leads to
a qualitative effect which is completely beyond the quasiclassical
theory.  It is well-known that the gauge-invariant potential $\Phi $
does not enter the equation for the quasiclassical stationary retarded
Green's function, and therefore, the spectrum found from then
Eilenberger equation is completely unaware of its presence.  However,
the potential $\Phi $ changes the actual value of the Fermi energy at
which the Cooper pair condense and, therefore, modifies the value of
the Fermi momentum by $\delta p_{F} = \Phi / v_{F}$. Varying $\Phi $,
one controls local value $p_{F}$ and the phase acquired by the
particle on the intefering paths. Proper choice of conditions, may
turn the constructive interference into destructive, and, therefore
modify the loop contribution to the density of states. Thus, the
loop-like paths provide a physical mechanism for influencing the
density of states in a superconductor by the non-equilibrium potential
$\Phi $.

\begin{acknowledgments}
This work was supported by the University of   Ume{\aa}.
\end{acknowledgments}

\appendix

\section{Observables}\label{obser}

For calculating observables the Green's function at coinciding
arguments $G(x,x)$ is needed. It follows directly from the continuity
of $G(x,x')$ for $x=x'$ and the definition of $\hat{g}^{(\pm)}$ that
\begin{eqnarray}
\hat{G}(x,x) &=& 
\frac{1}{2} \left( \hat{G}(+x,x) +  \hat{G}(-x,x) \right) =
\nonumber \\
&=& 
\frac{1}{2} \left( \hat{G}^{(+)} (+x,x) +  \hat{G}^{(-)} (+x,x)
+  \hat{G}^{(+)} (-x,x) +  \hat{G}^{(-)} (-x,x) \right) = \nonumber \\
&=&\frac{1}{2} \left( \hat{g}^{(+)} (x) + \hat{g}^{(-)}(x) \right) \; .
\nonumber 
\end{eqnarray}
The local density of states can be written
in terms of $\hat{\mathbb G}$ as follows
\[
\nu (x) = \frac{i}{2\pi} \lim \limits_{x \rightarrow x'} \mbox{Tr} \;
\left( {\mathbb G}(x,x') - {\mathbb G}^{A}(x,x') \right) \; . 
\]
From here the following expression for LDOS
via the 1-point Green's function $g^{(\pm)} (x)$ can be found
\[
\nu ( x ) = 
\frac{1}{2\pi v_{F}} \Re \;
 \mbox{Tr} \; 
 \left[ \Big( \hat{g}^{(+)} (x) + \hat{g}^{(-)}(x) \Big) \hat{\tau}_{z} \right]
\]
This is the formula for the local density of states at a point $x$
expressed in terms of the 1-point Green's function $g(x)$.

The standard quantum mechanical expression for the current density via
Gor'kov Green's function reads
\[
j(x) = \frac{1}{4m} \lim \limits_{x \rightarrow x'} \left(
\frac{\partial}{\partial x} - \frac{\partial}{\partial x'} \right) \;
\mbox{Tr} \hat{\mathbb G}^{K} (x,x') \; .
\]
In the equilibrium the Keldysh Green's function can be expressed as
$\hat{\mathbb G}^{K} = (\hat{\mathbb G} - \hat{\mathbb G}^{A}) \tanh
\varepsilon/2T$.  The differentiation with respect to $x$ and $x'$ is
performed using identities $\partial_{x} \hat{G}^{(\pm)} = \pm i p_{G}
\hat{G}^{(\pm)}$ and $\partial_{x'} \hat{G}^{(\pm)} = \mp i p_{G}
\hat{G}^{(\pm)}$. This leads to
\[
j(x) = -v_{F} \tanh \frac{\varepsilon}{2T} \;
\lim \limits_{x \rightarrow x'}
 \Im \;
\frac{1}{i v_{F}} \; \mbox{Tr} \left( \hat{G}^{(+)} - \hat{G}^{(-)}
\right) \tau_{z} \; .
\]
Finally we take the limits $x \rightarrow x'$ and get the expression
for the current density in terms of 1-point Green's function $\hat{g}(x)$
\[
j(x) = \frac{1}{2} \tanh \frac{\varepsilon}{2T} \; \Re \; \mbox{Tr} 
\left[ \Big( \hat{g}^{(+)}(x) - \hat{g}^{(-)} (x) \Big) \hat{\tau}_{z}
\right] \; .
\]

\section{Evolution matrix}\label{U}

The purpose of this section is to present a method for calculation the
quasiclassical evolution matrix.

One can always present $U(x,x')$ in the following form
\begin{equation}
 U(x,x') = 
a(x,x') {1 \choose \alpha (x,x')}\otimes \left(1,0 \right)  
+ b(x,x') {\beta (x,x') \choose 1} \otimes \left(0,1 \right)
\label{9bc}
\end{equation}
where 
\begin{equation}
a(x,x)=b(x,x) =1 \quad,\quad  \alpha (x,x)= \beta (x,x)=0
\label{acc}
\end{equation}
as required by the condition $U(x,x)=1$.  Since the evolution matrix
must generate a solution for arbitrary initial condition at $x'$, the
two terms in Eq.~(\ref{9bc}) must satisfy the quasiclassical equation
separately.

From Eq.~(\ref{pfc}).  the two-component ``wave function'' ${u \choose
v}= {a \choose a \alpha } $ and $={b \beta \choose b}$ considered as
functions of $x$, satisfy the Andreev-type linear first-order
differential equation.  It is known from the literature
\cite{Nag98,SchMak95} that the ratio $u/v$ obeys the nonlinear Riccati
equation,
\[
i {\partial\over{\partial x}} \alpha  =
2\varepsilon\alpha  + \Delta ^{*R}
+ \Delta\alpha^{2} 
\quad,\quad 
-i {\partial\over{\partial x}} \beta   =
2\varepsilon\beta   + \Delta ^{*R} \beta^2
+ \Delta\;.
\]  
with the initial condition in Eq.~(\ref{acc}).

Known $\alpha $ and $\beta $, parameters $a(x,x')$ and $b(x,x')$ can
be found as\cite{SheOza00}
\[
a(x,x')=  e^{
i\int\limits_{x'}^{x} dx\left(\varepsilon
 + \Delta\alpha\right)
}
\quad,\quad 
b(x,x')=  e^{
-i\int\limits_{x'}^{x} dx\left(\varepsilon
 + \Delta^{*R}\beta \right)
}
\]
This completes the derivation of the evolution matrix in a general
case.


\begin{thebibliography}{99}
\bibitem{gA} 
The identification of ``+''(``-'') with the electron (hole), is
the opposite when one analyses the advanced Green's function
$\hat{g}^{A}$.
\end{thebibliography}
\end{document}